\documentclass[format=acmsmall, review=false, screen=true]{acmart}

\usepackage{url}
\usepackage{caption}
\usepackage{graphicx}
\usepackage{float} 
\usepackage{subfigure}
\usepackage{subcaption}
\usepackage{balance}

\usepackage{pgfplots}

\acmJournal{TOSEM}
\usepackage{algorithm}
\usepackage{algorithmic}
\usepackage{wrapfig}
\usepackage{multirow}
\usepackage{makecell}
\usepackage{utfsym}
\usepackage{listings}
\usepackage{amssymb}
\usepackage{xcolor}
\usepackage[table]{xcolor}
\usepackage{pgf}
\usepackage{colortbl}
\definecolor{codegreen}{rgb}{0,0.6,0}
\definecolor{codegray}{rgb}{0.5,0.5,0.5}
\definecolor{codeorange}{rgb}{1,0.49,0}
\definecolor{backcolor}{rgb}{0.95,0.95,0.96}

\usepackage{tcolorbox}
\usepackage{colortbl}
\usepackage{framed} 
\lstdefinestyle{mystyle}{
 backgroundcolor=\color{backcolor},
 commentstyle=\color{codegray},
 keywordstyle=\color{codeorange},
 numberstyle=\tiny\color{codegray},
 stringstyle=\color{codegreen},
 basicstyle=\ttfamily\footnotesize,
 breakatwhitespace=false,
 breaklines=true,
 captionpos=b,
 keepspaces=true,
 numbers=left,
 numbersep=5pt,
 showspaces=false,
 showstringspaces=false,
 showtabs=false,
 tabsize=2,
 xleftmargin=10pt,
}
\begin{document}

\title{Bug Priority Change Prediction: An Exploratory Study on Apache Software}
\author{Guangzong Cai}
\affiliation{%
  \institution{Central China Normal University}
  \country{China}
}
\email{guangzongcai@mails.ccnu.edu.cn}

\author{Zengyang Li}
\authornote{Corresponding author.}
\affiliation{%
  \institution{Central China Normal University}
  \country{China}
}
\email{zengyangli@ccnu.edu.cn}

\author{Peng Liang}
\affiliation{%
  \institution{Wuhan University}
  \country{China}
}
\email{liangp@whu.edu.cn}

\author{Ran Mo}
\affiliation{%
  \institution{Central China Normal University}
  \country{China}
}
\email{moran@ccnu.edu.cn}

\author{Hui Liu}
\affiliation{%
  \institution{Huazhong University of Science and Technology}
  \country{China}
}
\email{hliu@whu.edu.cn}

\author{Yutao Ma}
\affiliation{%
  \institution{Central China Normal University}
  \country{China}
}
\email{ytma@ccnu.edu.cn}

\renewcommand{\shortauthors}{Cai et al.}

\begin{abstract}

Bug fixing is a critical activity in the software development process. In issue tracking systems such as JIRA, each bug report is assigned a priority level to indicate the urgency and importance level of the bug. The priority may change during the bug fixing process, indicating that the urgency and importance level of the bug will change with the bug fixing. However, manually evaluating priority changes for bugs is a tedious process that heavily relies on the subjective judgment of developers and project managers, leading to incorrect priority changes and thus hindering timely bug fixes.
Given the lack of research on bug priority change prediction, we propose a novel two-phase bug report priority change prediction method based on bug fixing evolution features and class imbalance handling strategy. Specifically, we divided the bug lifecycle into two phases: bug reporting and bug fixing, and constructed bug priority change prediction models for each phase. To evaluate the performance of our method, we conducted experiments on a bug dataset constructed from 32 non-trivial Apache projects. The experimental results show that our proposed bug fixing evolution features and the adopted class imbalance handling strategy can effectively improve the performance of prediction models. The F1-score of the prediction model constructed for the bug reporting phase reached 0.798, while the F1-weighted and F1-macro of the prediction model constructed for the bug fixing phase were 0.712 and 0.613, respectively. Furthermore, we explored the cross-project applicability of our prediction models and their performance at different priority levels. The findings indicate large variations in model performance across different projects, although the overall scores remain decent. Meanwhile, the predictive performance across various priority levels remained relatively consistently high. 
\end{abstract}




\maketitle

\section{Introduction}
\begin{figure*}
    \centering{\includegraphics[width=4.0in]{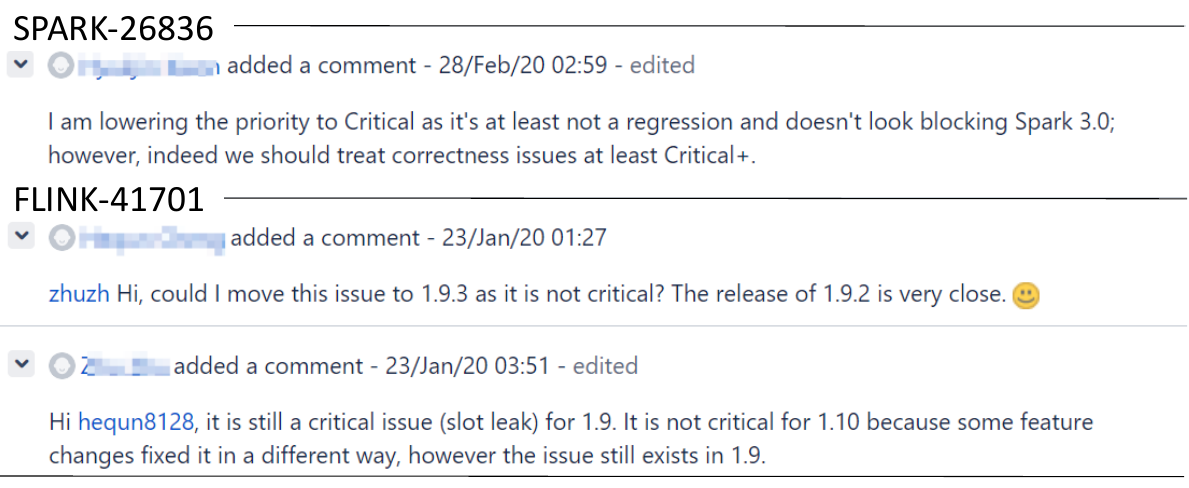}}
    \caption{Comments related to priority changes in bugs \href{https://issues.apache.org/jira/browse/SPARK-26836}{SPARK-26836} and \href{https://issues.apache.org/jira/browse/FLINK-14701}{FLINK-14701}.}
    \label{fig:2bug}
\end{figure*}

In the modern software development process, using an issue tracking system (ITS), such as JIRA, to manage and track software bugs has become a standard practice \cite{FiKoLu2013}. Each bug report is assigned a priority level to signify the urgency and importance of its resolution \cite{GoDaCo2019, RaUmYiZhIl2019, NoZhWaZo2019, LaDeSoVe2011}. As bug fixing progresses, the priority of a bug may change, reflecting the dynamic changes in the urgency and importance of the bug. Although the ITS provides powerful bug management capabilities, it still faces problems in practice. 
One significant problem is the time-consuming nature of manually evaluating and adjusting bug priorities \cite{IqNaJaAlYaAl2020, UdGhDeNaSh2017,UmLiSu2018}, a process that can be particularly cumbersome in large-scale projects characterized by a high volume of bugs and frequent priority changes \cite{BuFaKrPePlSu2023}. Another key problem is the subjectivity involved in determining priority changes, which often relies on the judgment of developers or project managers \cite{ZoLoChXiFeXu2018}. This subjective approach can lead to inconsistencies and a lack of objectivity, potentially resulting in critical issues not being addressed in a timely manner \cite{SaHaHa2020}. 
For instance, participants might not align their priorities with established community guidelines, which can lead to questioning and communication barriers among team members. Even when efforts are made to adjust priorities logically, participants may experience confusion due to uncertainty about the appropriate priority levels. Priority changes are frequently intertwined with software version updates, and differing opinions among participants regarding the same bug can lead to misjudgments about the bug priority. These factors not only diminish team efficiency but also risk preventing critical problems from receiving timely attention and resolution. Specifically, Fig.~\ref{fig:2bug} illustrates several comments related to priority changes in two bugs (SPARK-26836 and FLINK-14701). In bug SPARK-26836, a participant explicitly noted uncertainty while downgrading the bug priority from Blocker to Critical. In bug FLINK-14701, conflicting views on bug priority arose due to differing interpretations of project version iterations. Collectively, these examples highlight the subjective and complex nature of priority adjustments in practice.
Predicting changes in bug priority holds substantial practical value. It enables project managers and developers to anticipate which bugs might escalate in urgency, thereby facilitating proactive resource allocation and scheduling \cite{SiMiSh2017, MeMa2008, MaChMoHeUd2024}. In addition, accurate prediction of priority changes supports the optimization of development process, ensuring that pivotal issues are resolved promptly, which contributes to improved software product quality.

However, there is a notable gap in research on the prediction of bug priority changes. Most of the existing literature has concentrated on predicting static bug priorities. Although numerous studies have employed machine learning (ML) and deep learning (DL) techniques to build models for predicting bug priorities, they have largely overlooked the dynamic nature of priority changes during the resolution process \cite{IzAkHe2022}. Furthermore, several weaknesses remain within the existing bug priority prediction approaches \cite{HeYaPaHuZh2023}.
First, the categorization of predicted outcomes often lacks the necessary granularity \cite{ShAl2022}, rendering the predictions insufficiently informative for guiding the bug-fixing workflow. Second, many studies have focused exclusively on extracting basic or textual features from bug reports, failing to leverage the full spectrum of information contained within them \cite{GoSo2021, AkCaBe2018, KuSi2020}. In fact, bug fixing is a dynamic process, and its dynamically changing characteristics are crucial \cite{IzAkHe2022}.
Third, the class imbalance problem \cite{XiLoShWaYa2015} leads to poor performance of prediction models on a few classes \cite{YaRa2024}. Finally, the approach taken to define the prediction task also varies: while some studies treat it as a multi-class classification problem, others frame it as a regression task \cite{TiAlLoHa2016}. The former may disregard the ordinal relationship between priority levels, whereas the latter could introduce additional estimation errors. 
Additionally, predicting bug priority changes also faces unique challenges. First, it involves modeling the dynamic evolution of bug reports over time, requiring temporal features that capture how bugs are discussed, reassessed, and modified during their lifecycle. Second, existing models for bug priority prediction generally assume a single prediction point (i.e., at bug reporting), whereas bug priority changes can occur multiple times and at different phases. Third, most bug priority prediction models treat the task as a classification task based on static features, often ignoring the influence of project context, developer behavior, and bug filed change history. These aspects are crucial for predicting whether and how a bug's priority changes. Therefore, directly applying bug priority prediction methods to the task of bug priority change prediction may lead to suboptimal or misleading results. 
In summary, current methodologies for bug priority prediction exhibit substantial limitations, especially in coping with the dynamic nature of bug priorities. In light of the scarcity of research on predicting priority changes and the weaknesses identified in existing bug priority prediction approaches, there is a pressing need to develop a more effective method for predicting these dynamic changes in bug priority.

\begin{figure*}
    \centering{\includegraphics[width=5.0in]{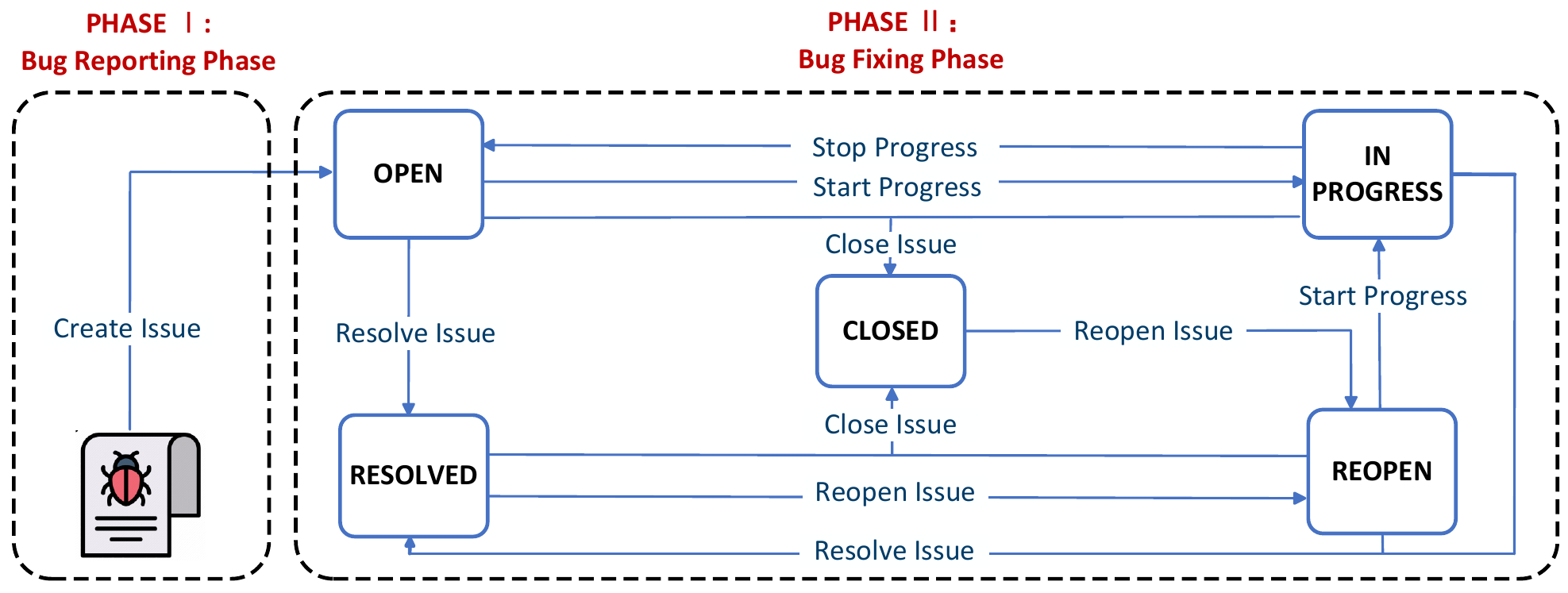}}
    \caption{Bug report lifecycle.}
    \label{fig:bugPhase}
\end{figure*}

Therefore, in this work, we aim to construct a bug priority change prediction method. It is worth emphasizing that ``\textbf{bug priority change prediction}'' and ``\textbf{bug priority prediction}'' are two completely \textbf{different tasks}. Bug priority change prediction aims to identify whether a bug will undergo priority adjustment and its changing trend during its lifecycle, while bug priority prediction focuses on determining the initial priority of a bug based on the information when the bug report is first submitted. This study specifically concentrates on priority change prediction, that is, analyzing and predicting the possible priority changes of bugs during their fixing process. 
As shown in Fig. \ref{fig:bugPhase}, the bug lifecycle is divided into two phases for priority change prediction: bug reporting phase (Phase I) and bug fixing phase (Phase II). Through constructing the prediction models for these two phases, this work can cover the entire lifecycle of software bugs, enabling practitioners to manage priority changes more reasonably and efficiently. These two phases are detailed as follows: 

    \textbf{Phase I: Bug Reporting Phase.}
This phase refers to the process of reporting a bug on the JIRA platform. At this phase, the prediction model proposed in our study can predict whether the priority of a bug will change in the subsequent fixing process after it is reported and assigned an initial priority. This prediction helps with early resource optimization and ensures timely attention to key issues. 
It is worth noting that even if the model predicts that the priority of a bug will change in the subsequent process, it does not necessarily imply that the initial priority assignment is incorrect. The project development and bug fixing processes are inherently dynamic and iterative, and the initial priority may be based on the best judgment made with the information and environment available at that time. As the project evolves and the environment changes, the previously appropriate priority of the bug may no longer be applicable, thus, adjustment of its priority needs to be made. 

\textbf{Phase II: Bug Fixing Phase.}
This phase encompasses the entire process from bug reporting to final resolution, including state transitions such as ``OPEN'' to ``IN PROCESS'', ``RESOLVED'', ``CLOSED'', or ``REOPENED''. Throughout this process, the priority of bugs may be changed, and practitioners may change priorities based on project progress or work arrangements. The prediction model proposed in our study is used to predict the specific changes in bug priority during this phase, which aids in the dynamic adjustment of the team's bug-fixing plan. 
Specifically, when practitioners consider adjusting bug priorities, the model can predict the new level after the priority change based on the current bug fixing evolution information (e.g., project iterations, bug reporter experience, bug comment changes, etc.). By referring to the model's predictions, practitioners can make more informed and reasonable decisions regarding priority changes. This is particularly valuable in situations where incomplete information or subjective judgments might otherwise lead to unexpected priority assignments. 


We proposed a two-phase bug report priority change prediction method based on bug fixing evolution features and class imbalance handling strategy. Firstly, we collected and filtered bug report data from 32 non-trivial Apache projects in JIRA. 
Secondly, we constructed bug fixing evolution features with respect to the project, the reporter, the comments, and the historical change items of the bug based on this dataset.  
Thirdly, we designed a comprehensive class imbalance handling strategy, comprised of a conditional mixed sampling method at the data level as well as a combination of the soft voting ensemble method and the cost-sensitive learning method at the algorithm level. 
Finally, our method constructs different prediction models for the two phases. 1) In Phase I, we used four ML algorithms—Random Forests (RF), K-Nearest Neighbors (KNN), Support Vector Machine (SVM), and Extreme Gradient Boosting (XGBoost)—to construct a prediction model. This model is designed to predict whether a newly reported bug will undergo a priority change during the subsequent bug fixing process; 2) In Phase II, we use an artificial neural network to build the other prediction model. This model is specifically aimed at predicting the new priority level to which a bug's priority will change if a change is predicted.
Admittedly, relying solely on Phase II is not enough, because Phase II does not cover the early phases of the bug lifecycle. In contrast, Phase I can provide an important early warning signal at the time point when the bug report is submitted, which is crucial for dynamically adjusting project management and resource allocation.
In practice, it can be beneficial for project management teams to identify potential risks at the time point when the bug report is submitted so that they can formulate response strategies in a timely manner. The early warning provided by Phase I just meets this need, allowing the team to make predictions and preparations in the early phase of the bug lifecycle.
Combining Phase I and Phase II can form a comprehensive prediction system, providing forward-looking change warnings and real-time priority adjustment suggestions in different phases of the bug lifecycle.

The main \textbf{contributions} of this work are summarized as follows: 
\begin{itemize}
    \item \textbf{Pioneering the prediction of bug priority changes.} To our knowledge, this is the first study to move beyond static bug priority prediction and address the dynamic nature of bug priorities. We shift the research focus from asking ``What is the initial priority?'' to ``Will the priority change, and how?''. This opens up a new and practical research direction focused on proactive, adaptive issue management.

    \item \textbf{A novel two-phase prediction method with a comprehensive class imbalance handling strategy.} We proposed a method that uniquely mirrors the bug lifecycle, with distinct models for the ``bug reporting phase'' and the ``bug fixing phase''. This approach is promising because it provides actionable insights at different phases: an \textbf{early warning} at the time of bug reporting (Phase I) and real-time \textbf{adjustment suggestions} during the fixing process (Phase II). This methodological contribution is supported by a carefully designed class imbalance strategy for each phase, which is critical for achieving robust performance on real-world, imbalanced data.

    \item \textbf{A new set of bug fixing evolution features that capture dynamic project context.} Existing studies predominantly rely on static features from the initial bug reports. Our key contribution is the design and validation of novel \textbf{evolution features} derived from four perspectives: project evolution, bug reporter history, comment dynamics, and historical change items. These features are crucial because they characterize the dynamic changes in a bug's characteristics over time, providing a much richer and more accurate basis for prediction than static snapshots. Our ablation studies confirm that these features significantly enhance model performance.
\end{itemize}

The remainder of this paper is organized as follows to address the identified research gaps and validate our contributions. \textbf{Section \ref{sec_relatedwork}} reviews the related work on bug priority prediction, highlighting the critical research gap: the lack of focus on the dynamic changes of bug priority, which this study aims to fill. To address this gap, \textbf{Section \ref{sec_RQ}} details our core contribution - the two-phase bug priority change prediction method, including the construction of dynamic bug fixing evolution features. \textbf{Section \ref{sec_evaluation}} is dedicated to rigorously evaluating our method. Through a series of ablation and comparative experiments, we demonstrated the effectiveness of our proposed features and class imbalance handling strategies. \textbf{Section \ref{sec_discussion}} interprets the study results and discusses their implications for practitioners and researchers. \textbf{Section \ref{sec_validity}} clarifies the validity threats of our study, examining the reliability of our data collection and the validity of our experimental results. \textbf{Section \ref{sec_conclusion}} concludes the paper and outlines promising directions for future research.

\section{Related Work}\label{sec_relatedwork}
Due to the lack of previous research on bug priority change prediction, we examined the related work in two aspects, i.e., bug priority change and bug priority prediction.

\subsection{Bug Priority Change}
To our knowledge, there are currently only two studies on bug priority changes. Almhana et al. revealed the reasons for the bug priority change from the perspective of practitioners \cite{AlFeKeSh2020}. Specifically, they interviewed practitioners and conducted quantitative analysis, summarizing the reasons for the bug priority change as follows: 1) the dependency of another bug’s fix incorrect priority; 2) type/domain of project; 3) category of the bug report; 4) lack of of time/heavy workload/tight schedule; 5) accident; 6) hot-fix request; 7) business requirements. Li et al. conducted a large-scale empirical study on bug priority changes in Apache open source software projects \cite{LiCaYuLiMoLi2024}. Their research shows that around 8\% of bugs undergo priority changes, with about 88\% of bugs undergo only one priority change, and most priority changes tend to be adjusted towards adjacent priority levels. The time distribution of priority changes indicates that the majority of changes occur within a few days after the bug report, and over half of the priority changes occur before the bug begins to be processed. Further analysis revealed that bugs with higher code change complexity or more comments are more likely to undergo priority changes. Meanwhile, bugs reported or assigned by specific participants are more likely to have their priority adjusted. These two studies uncovered the reasons behind priority changes as well as the probabilities and patterns associated with these changes, providing  insights for our feature selection process and class imbalance handling.

While these two foundational studies provide invaluable insights into the nature of bug priority changes \cite{AlFeKeSh2020, LiCaYuLiMoLi2024}, it is crucial to note that their contributions are primarily descriptive. They effectively answer ``why a bug's priority change happens'' and ``what happens'' by identifying the reasons for and statistical patterns of priority adjustments. However, a significant gap remains: they do not offer a predictive method to forecast whether and how a bug's priority will change in the future. Our current research directly addresses this gap. Building upon the understanding provided by these two prior studies, we proposed a method dedicated to predicting bug priority changes, moving from a descriptive understanding to a proactive and predictive capability of bug priority changes.

\subsection{Bug Priority Prediction}

ML and DL techniques have been extensively leveraged in the realm of software bug priority prediction. Various studies have employed a range of ML algorithms, including Support Vector Machine (SVM), Naive Bayes (NV), Decision Trees (DT), Random Forest (RF), K-Nearest Neighbors (KNN), Logistic Regression (LR), and AdaBoost, to tackle the problem of bug priority prediction \cite{KaMa2012, AlBa2013, ShAl2022, DhVeCh2020, MaAhHe2023}. The strength of these models lies in their ability to learn from structured data, but they often require extensive feature engineering and may struggle to capture the complex semantic nuances within bug reports. These investigations highlight the variance in model performance attributable to differences in datasets or features used, often through comparative analysis of predictive efficacy among distinct models. Our study similarly adopted a diverse array of ML techniques to build predictive models; however, we opted for an ensemble learning approach to combine multiple models rather than individually assessing each model's performance. 
Additionally, neural network-based methodologies, such as those involving Bert, Convolutional Neural Networks (CNN), and Graph Convolutional Networks (GCN), have demonstrated utility in predicting bug priorities \cite{KuSi2020, AlXiUmOs2024, YaRa2024, IzAkHe2022, UmLiIl2019, FaTaZhXuLi2021, LiChHuWaWaWaWa2022, AcGi2024}. Research findings suggest that these approaches have led to notable enhancements in prediction accuracy. For example, CNNs and BERT are powerful for processing textual descriptions in bug reports, but might not inherently capture the evolving, dynamic nature of a bug's lifecycle. GCNs can model relationships between bugs, but might overlook temporal features. While these approaches have led to notable enhancements in prediction accuracy, they still primarily focus on predicting a static priority at the time point of bug reporting. In our study, we propose to construct an artificial neural network model aimed at achieving effective prediction of bug priority changes.

Regarding feature extraction, research in bug priority prediction underscores the significance of incorporating diverse features like fixing time, textual descriptions, reporters, related reports, severity, and product types. The Summary, Comment, and Description sections within bug reports contain rich textual information, making the extraction of these textual features a pivotal component in prediction tasks \cite{AgGo2021, PaNa2024, HuShFaYuYaZh2022, WuLiLiZhZh2024, HuShFaYuYaZh2024}. Adopting a similar approach, our study utilizes RoBERTa to extract textual features from the Summary and Comment sections.
Addressing class imbalance is essential for ensuring robust model performance. Prior research has introduced measures such as cost-sensitive K-nearest neighbor methods or the automatic adjustment of imbalanced decision boundaries to mitigate this challenge \cite{SaHaHa2020, XiLoShWaYa2015, SuHuChQuMe2022, DaYa2023}. Consequently, our study implements several measures to alleviate the problem of class imbalance.

Cross-project validation is imperative to evaluate the practical applicability and effectiveness of the bug priority prediction model. Sharma et al. conducted training on one project and tested the model's generalization capability on another \cite{ShBeChSi2012}. In line with this approach, our study trains models across multiple projects and performs predictions on another project to gauge the model's true predictive power.
Specialized research efforts have concentrated on predicting the highest priority levels, such as the ``Blocker'' level in JIRA \cite{ChLiGuXuZh2020, XiLoShWaYa2015}. In contrast, our study aims to encompass change prediction across all priority levels.
It is noteworthy that defect fields can differ across various issue tracking systems. For instance, JIRA employs the ``Priority'' field to denote the importance and urgency of bugs, whereas Bugzilla uses ``Priority'' to indicate the timeframe for resolution and ``Severity'' to reflect the impact on users \cite{ZhChYaLeLu2016, TaXuWaZhXuLu2020, LiWaChJi2018, ZhYaLeCh2015}. Despite these differences, the methodologies adopted in these studies provide valuable insights applicable to our research.

It is worth noting that our study focuses on predicting changes to bug priority during a bug's lifecycle, rather than assigning an initial priority label at the time of bug reporting. Due to the substantial differences in problem formulation, target outcomes, and required features, it is not appropriate to directly compare our priority change prediction method with existing methods for initial priority prediction.

Nevertheless, reviewing prior studies on bug priority prediction remains crucial, as many of the machine learning and deep learning techniques employed in those studies influence our feature engineering and bug priority change prediction model design. Table~\ref{table:ml_dl_comparison} summarizes representative techniques from previous studies, outlining the types of features they utilize with their strengths and limitations. This comparative summary highlights two gaps in the literature - namely, the lack of characterization of the dynamic evolution of bugs and insufficient attention to the class imbalance problem - all of which motivate the design of our two-phase method that integrates both static and dynamic bug report features across the the bug's lifecycle.

\begin{table}[]
\centering
\caption{Comparison with Prior Bug Priority Prediction Studies}
\label{table:ml_dl_comparison}
\footnotesize
\begin{tabular}{|p{0.1\linewidth}|p{0.14\linewidth}|p{0.2\linewidth}|p{0.19\linewidth}|p{0.23\linewidth}|}
\hline
\textbf{Study / Approach} & \textbf{Technique Type} & \textbf{Feature Types Used} & \textbf{Strengths} & \textbf{Limitations} \\ \hline
\cite{KaMa2012, AlBa2013, ShAl2022, DhVeCh2020, MaAhHe2023} & 
SVM, RF, KNN, LR, NB, DT, AdaBoost. & Static + meta attributes (e.g., attachment, reporter, component). & Interpretable, effective for small-to-medium data, fast training. & Requires manual feature engineering; weak in handling textual and temporal dynamics. \\ \hline
\cite{AcGi2024, KuSi2020, LiChHuWaWaWaWa2022, UmLiIl2019, YaRa2024} & CNN, GCN. & Bug description + summary text. & Captures semantic structure of text; improved accuracy. & Requires large labeled datasets; no modeling of temporal behavior. \\ \hline
\cite{AlXiUmOs2024, IzAkHe2022} & BERT / RoBERTa-based fine-tuning. & Pretrained contextual embeddings from bug reports. & Strong NLP capabilities; no handcrafted features needed. & High resource consumption; lacks modeling of priority evolution over time. \\ \hline
\cite{AgGo2021, PaNa2024, WuLiLiZhZh2024, HuShFaYuYaZh2022} & Textual feature extraction + ML classifiers. & Summary, comments, description text (TF-IDF, embeddings). & Combines textual richness with classification models. & May not capture deeper semantics; ignores bug lifecycle dynamics. \\ \hline
\cite{SaHaHa2020, XiLoShWaYa2015, SuHuChQuMe2022, DaYa2023} & Cost-sensitive ML; Imbalance-aware KNN. & Class weight tuning; decision boundary reshaping. & Improves performance for minority classes. & Requires tuning; not directly tied to temporal modeling. \\ \hline
\cite{ShBeChSi2012} & Cross-project prediction. & Generic ML pipeline with inter-project evaluation. & Validates model generalizability across domains. & Performance drops in heterogeneous project settings. \\ \hline
\textbf{Our Study} & Ensemble ML + temporal modeling (2-phase). & Static + textual + evolutionary features. & Models lifecycle dynamics and priority evolution. & More complex modeling and data preprocessing required. \\ \hline
\end{tabular}
\end{table}

\subsection{Comparison between Bug Priority Prediction and Bug Priority Change Prediction}
To clarify the challenges addressed in this work, it is essential to provide a detailed comparison between traditional bug priority prediction and the task of bug priority change prediction. While both tasks relate to bug priority, they represent fundamentally different research problems in terms of their goals, temporal nature, and technical requirements. Existing methods for bug priority prediction are insufficient for bug priority change prediction because they are designed to analyze a static snapshot of a bug at the time of its creation. These methods cannot capture the dynamic, evolving signals that are the very triggers for a future priority change. We summarize the key distinctions in Table \ref{table:bp_comparison}.

\begin{table}[]
\centering
\caption{Comparison between Bug Priority Prediction and Bug Priority Change Prediction}
\label{table:bp_comparison}
\footnotesize
\begin{tabular}{|p{0.2\columnwidth}|p{0.34\columnwidth}|p{0.37\columnwidth}|}
\hline
\textbf{Dimension} & \textbf{Bug Priority Prediction} & \textbf{Bug Priority Change Prediction} \\ \hline
\textbf{Objective} & To assign the \textbf{correct initial priority} to a newly submitted bug report. & To predict if and how a bug's priority will be \textbf{modified in the future} during its lifecycle. \\ \hline
\textbf{Problem Type} & A \textbf{static, one-time classification} of a snapshot of information at the time a bug is reported. & A \textbf{dynamic, event-based prediction} problem that models a state transition over time. \\ \hline
\textbf{Feature Requirements} & Relies on \textbf{static attributes} from the initial report (e.g., summary, description, component, reporter). & Requires \textbf{dynamic, evolutionary features} that capture changes over time (e.g., comments, project workload changes, history of field modifications). \\ \hline
\textbf{Key Challenges} & \textbf{Shared Challenges: }Feature selection, model building, and class imbalance processing. & \textbf{Shared Challenges: }Feature selection, model building, and class imbalance processing. \textbf{Unique Challenges: Rare Event Detection:} Priority changes are highly infrequent (8\% \cite{LiCaYuLiMoLi2024}), creating an extreme class imbalance problem. \textbf{Evolutionary Feature Engineering:} Requires creating novel features from raw event logs that signal changes. \\ \hline
\textbf{Practical Application} & At bug submission by triage teams to improve \textbf{initial classification efficiency}. & Throughout the bug lifecycle by project managers for \textbf{proactive risk management and dynamic planning}. \\ \hline
\end{tabular}
\end{table}

\section{Method}\label{sec_RQ}
\begin{figure*}
    \centering{\includegraphics[width=5.5in]{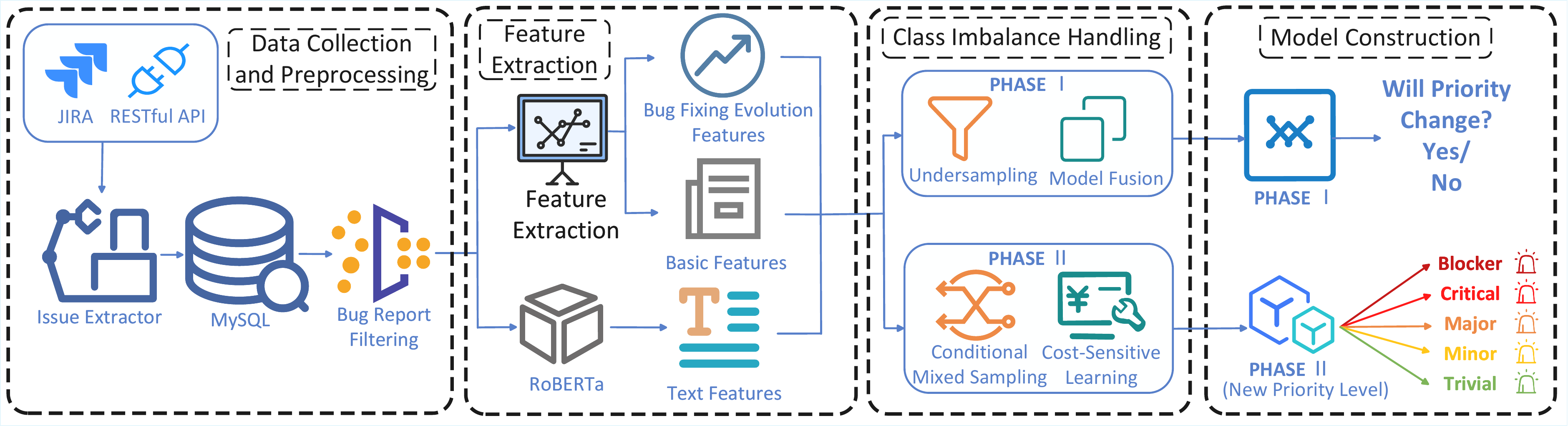}}
    \caption{The process of our proposed bug priority change prediction method.}
    \label{fig:PredMethod}
\end{figure*}

We propose a two-phase software bug report priority change prediction method based on Bug Fixing Evolution features and an Imbalanced Class handling strategy. The overall process of the prediction method is illustrated in Fig. \ref{fig:PredMethod}. Specifically, the method comprises four steps, which will be briefly introduced here and subsequently detailed in the following subsections:
\begin{enumerate}
\item \textbf{Data Collection and Preprocessing}: We obtained bug report data from 32 non-trivial Apache projects through the JIRA API. First, we carefully selected the representative projects from all Apache projects to ensure the richness and diversity of the dataset. Then, the collected data underwent a series of rigorous preliminary filtering and preprocessing steps to improve the quality and usability of the data. This includes retaining only bug reports with clearly marked priorities and converting priorities into numerical form for subsequent analysis. In addition, we also paid special attention to incorrect priority modifications caused by users' personal reasons (such as regret) and corrected them. Through the above measures, we ensured that the bug report data used is highly accurate and reliable.
\item \textbf{Feature Extraction}: We derived bug fixing evolution features from four perspectives: project evolution, bug reporter, comment changes, and historical change items, using part of bug report fields as basic features. Additionally, we employed the RoBERTa pre-trained model to extract text features from the Summary and Comment fields of the bug reports. When dealing with bug feature extraction, we paid special attention to the impact of chronological order on features. Since the features in the fixing process may change over time, after we extracted a bug, when parsing the priority change record of the bug, we restored the features of the bug to the state at the time of the priority change. For example, on September 15, 2022, at 15:00, we extracted a bug that had 10 comments. However, this bug underwent a priority change on June 20, 2020, at 18:00, and at that time, it only had 6 comments. To restore the bug to its state as of 18:00 on June 20, 2020, we deleted all activity records after that time, including the 4 subsequent comments, and similarly adjusted other fields. Through this way, we ensured that all fields accurately reflected the true state of the bug at 18:00 on June 20, 2020, thereby guaranteeing that the data at each time point precisely represented the actual condition of the bug at that moment.

\item \textbf{Class Imbalance Handling}: We designed a comprehensive class imbalance handling strategy at both the data and algorithm levels. In Phase I, we addressed class imbalance in the dataset through undersampling. Then, we employed four ML techniques and constructed a prediction model via model fusion. In Phase II, we tackled the class imbalance problem at the data level through conditional mixed sampling and constructed a prediction model at the algorithm level using DL techniques and cost-sensitive learning methods.
\item \textbf{Model Construction}: In Phase I, we constructed a binary prediction model capable of predicting whether bugs will undergo priority changes upon their initial reporting and assignment. This early prediction aids in optimizing resource allocation and prioritizing management, ensuring critical bugs are addressed promptly. In Phase II, we build a multi-class model to predict the new priority level to which the bug will be changed. This phase’s predictions help dynamically adjust the bug fixing plan throughout the process, ensuring efficient resource allocation and timely priority adjustments.
\end{enumerate}

\subsection{Process of Bug Collection and Preprocessing}
\label{process_data_processing}

Leveraging the RESTful API provided by JIRA\footnote{\href{https://developer.atlassian.com/cloud/jira/platform/rest/v3/intro/}{https://developer.atlassian.com/cloud/jira/platform/rest/v3/intro/}}, we developed a customized Issue Extractor to retrieve issue data from Apache projects and store them in a MySQL database. Specifically, we applied the following criteria to select projects:

\begin{itemize}\setlength{\itemsep}{0pt}\setlength{\parskip}{0pt}
  \item \textbf{C1:} The project adopts all five standard bug priority levels - \emph{Blocker}, \emph{Critical}, \emph{Major}, \emph{Minor}, and \emph{Trivial} - to label the priority of each issue. 
  \item \textbf{C2:} The project age exceeds five years, ensuring sufficient project history and maintenance continuity. 
  \item \textbf{C3:} The corresponding code repository contains more than 3,000 revisions (i.e., commits), reflecting sustained development activity. 
  \item \textbf{C4:} The project includes at least 150 bug reports with one or more priority-change events, ensuring that priority evolution is sufficiently observable.
\end{itemize}

These criteria combined to ensure the scale and representativeness of the selected projects, ultimately resulting in a selection of 32 non-trivial Apache projects that met these criteria.

\textbf{Unit of analysis and data status.}
Our study exclusively uses bug reports whose final status is ``Resolved'' or ``Closed'' at the time of data snapshot. This ensures that we train ML models on historical, complete priority trajectories, avoiding instability from open issues that may still change. In Phase~II, the unit of analysis is a single priority-change event (not an entire bug): for example, if one bug changes priority three times, it contributes three event samples.

\begin{figure*}
    \centering{\includegraphics[width=3.8in]{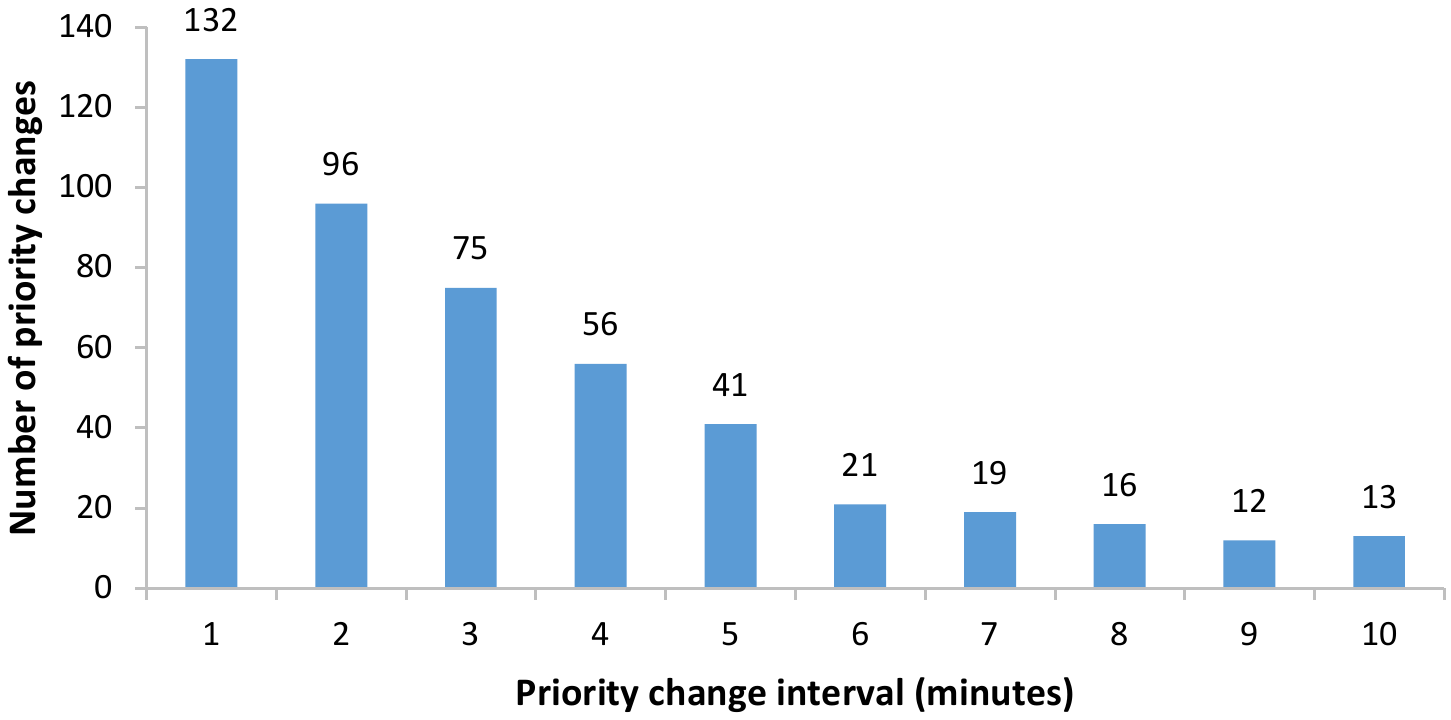}}
    \caption{Determining the 5-Minute threshold for priority change filtering.}
    \label{fig:5minutes}
\end{figure*}

\begin{figure*}
    \centering{\includegraphics[width=5.5in]{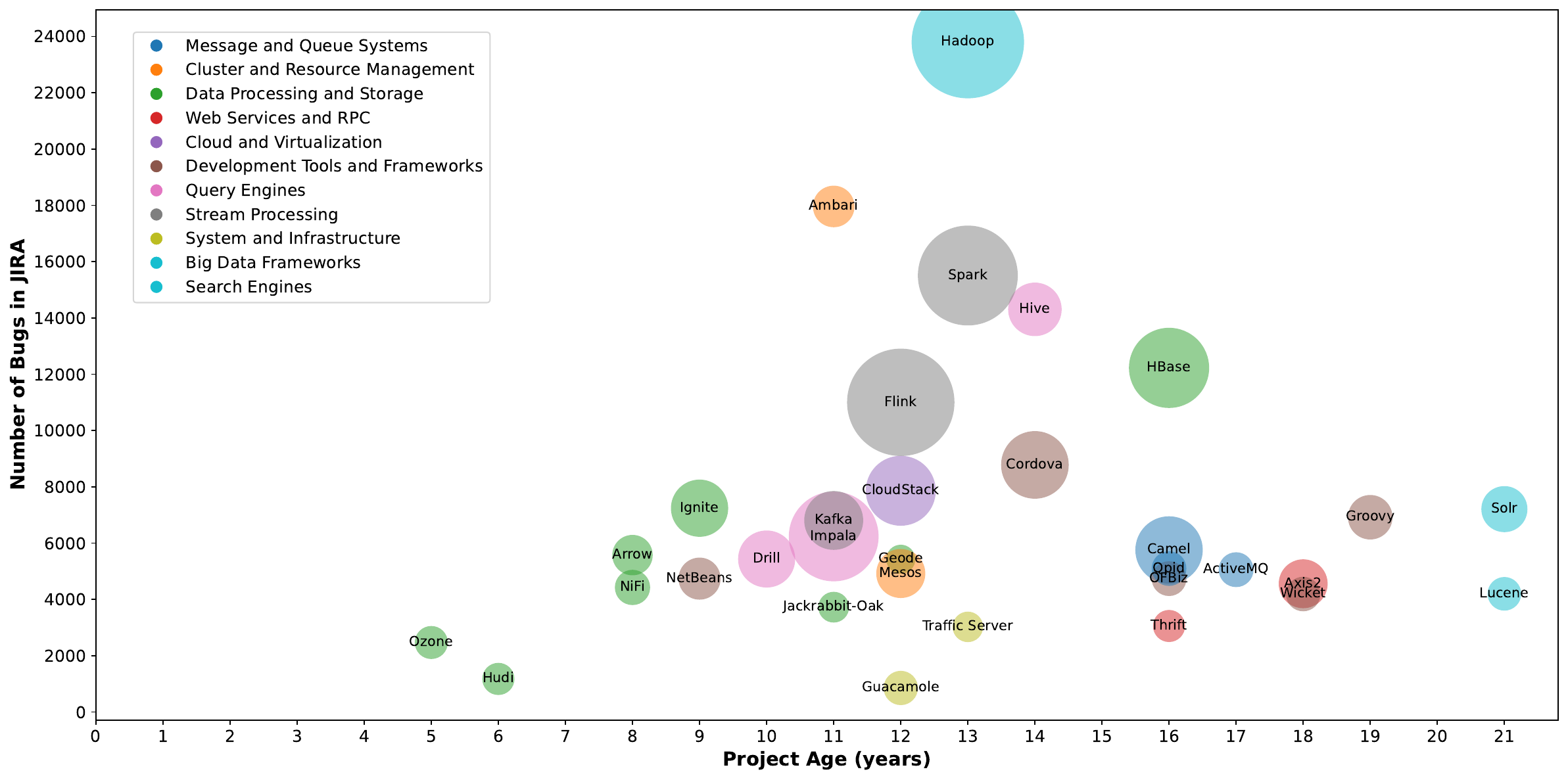}}
    \caption{Visualization of the diverse characteristics of the 32 selected Apache projects.}
    \label{fig:projects_bubble}
\end{figure*}

\textbf{Filtering and preprocessing pipeline.} We perform the following steps to construct the dataset and labels for both phases:

\begin{enumerate}
    \item \textbf{Project and issue retrieval \& filtering.}
    \begin{enumerate}
        \item \textbf{Project scope:} We focus on Apache open-source projects due to the availability of large, well-documented issue histories and development processes. The 32 selected projects are summarized in Fig. \ref{fig:projects_bubble}.
        \item \textbf{Issue type:} From each project, we query all issues with ``issuetype = Bug''.
        \item \textbf{Final status constraint:} We keep only bugs whose final status is ``Resolved'' or ``Closed''. Bugs that are still open at the snapshot time are excluded to avoid unstable labels.
        \item \textbf{Full history:} For every retained bug, we extract its complete changelogs (field-level histories), comments, and metadata (e.g., reporter, component, create/close time).
        \item \textbf{Sanity checks:} We remove malformed histories (e.g., missing timestamps in critical fields).
    \end{enumerate}
    \item \textbf{Priority field normalization (vocabulary \& encoding).}
    \begin{enumerate}
        \item \textbf{Canonical set:} We restrict priorities to the five canonical JIRA levels: ``Blocker'', ``Critical'', ``Major'', ``Minor'', and ``Trivial''. Records with non-canonical or deprecated values are either mapped to the closest canonical value or removed if mapping is ambiguous.
        \item \textbf{Standardization:} We normalize the priority field by converting all values to lowercase, trimming extraneous whitespace, and mapping various project-specific aliases (e.g., alternative names for the highest priority) to their corresponding canonical terms.
        \item \textbf{Numeric encoding (for modeling convenience):} We encode ``{Blocker''=1, ``Critical''=2, ``Major''=3, ``Minor''=4, and ``Trivial''=5.}
    \end{enumerate}
    \item \textbf{Check and correct multiple priority changes within a short period of time due to personal reasons.} To determine the time threshold for changing priorities due to personal reasons, we manually analyzed the cases of multiple modifications within a short period of time. Based on our empirical evaluation, we focused our analysis on records with a time interval of less than 10 minutes, a window sufficient to capture such short-term editing behavior while excluding unrelated, longer-term modifications. Specifically, we searched all bugs for: records with a time interval of less than 10 minutes between two consecutive priority changes; or records of priority changes within 10 minutes after the bug is reported. The results are shown in Fig. \ref{fig:5minutes}. The horizontal axis in the figure represents the time interval, and the vertical axis represents the number of priority changes. For example, there are 132 records with a time interval of less than 1 minute, 96 records with a time interval of 1-2 minutes. As shown in Fig. \ref{fig:5minutes}, the number of changes within 1 minute is the highest, and then gradually decreases; after 5 minutes, the number of changes within each minute tends to be stably distributed. Therefore, we set the time threshold to 5 minutes. The reason for not taking 1 minute is that the number of changes within 2-4 minutes is still large and fluctuates significantly. Based on this threshold, we formulated the following rules: a) If the bug reporter changes the priority within 5 minutes after the report, the initial priority is set to the priority last modified within these 5 minutes, and the change record is deleted. b) If the same person changes the priority multiple times and back to its original value within 5 minutes, all changes will not be counted. c) If the same person changes their priority twice within 5 minutes, they will be combined into one single record.
    \item \textbf{Priority-change event extraction.}
    From the normalized histories, we enumerate every priority change \(\langle P_{\text{old}}\rightarrow P_{\text{new}}\rangle\) with its timestamp \(t_c\). Each change constitutes one Phase~II training sample. For a bug with multiple changes, multiple samples are created (after applying the short-interval rules in Step~(3)). For every priority change, we also record contextual snapshots (e.g., latest comment signals, field states) strictly up to but excluding \(t_c\).
    \item \textbf{Label construction for Phase~I and Phase~II.}
    \begin{enumerate}
        \item \textbf{Phase~I (binary) label:} For each bug, \(y^{(1)}=1\) if the bug experienced at least one priority change during its lifecycle; otherwise \(y^{(1)}=0\). Phase~I features are computed using only information available at creation time.
        \item \textbf{Phase~II (multiclass) label:} For each extracted priority change, the target is the {new priority \(P_{\text{new}}\in\{\text{Blocker},\text{Critical},\text{Major},\text{Minor},\text{Trivial}\}\) (or its numeric encoding).}
    \end{enumerate}
    \item \textbf{Temporal leakage control.}
    To ensure a faithful historical setting, feature computation obeys strict temporal cut rules, which prevent label leakage and align training with real-world usage:
    \begin{enumerate}
        \item Phase~I uses only fields and counts available at bug report creation time.
        \item Phase~II uses only signals observed before the event time \(t_c\) (e.g., comment dynamics, recent field changes), excluding any future information. 
    \end{enumerate}
    \item \textbf{Project-level descriptive statistics and diversity evidence.}
    We compiled statistics for each project to demonstrate the richness and diversity of our dataset. Fig. \ref{fig:projects_bubble} visualizes the diverse characteristics of the 32 selected Apache projects. Each bubble represents a project, determined by its age (x-axis) and the total number of bugs in JIRA (y-axis). The size of the bubble is proportional to the number of bugs with priority changes, while the color indicates its primary application domain. This figure highlights the significant heterogeneity in our dataset. More detailed information about each project can be found in the reproducible package \cite{dataset}.
\end{enumerate}

Finally, we compute, for each project, the total number of bugs, the number of bugs with priority changes, and the probability of priority change, as shown in Table~\ref{table:proportion}. These descriptive statistics provide concrete evidence regarding dataset composition and help contextualize the results reported later.

\begin{table}[]
\centering
\caption{Proportion of bugs with priority changes.}
\label{table:proportion}
\scalebox{0.62}{
\begin{tabular}{lrrr|lrrr|lrrr}
\hline
\textbf{Project} & \multicolumn{1}{c}{\textbf{\#Bug in JIRA}} & \multicolumn{1}{c}{\textbf{\#BugPC}} & \multicolumn{1}{c|}{\textbf{\%}} & \textbf{Project} & \multicolumn{1}{c}{\textbf{\#Bug in JIRA}} & \multicolumn{1}{c}{\textbf{\#BugPC}} & \multicolumn{1}{c|}{\textbf{\%}} & \textbf{Project}      & \multicolumn{1}{c}{\textbf{\#Bug in JIRA}} & \multicolumn{1}{c}{\textbf{\#BugPC}} & \multicolumn{1}{c}{\textbf{\%}} \\ \hline
ActiveMQ         & 5,058                                      & 216                                  & 4.3\%                            & Guacamole        & 856                                        & 209                                  & 24.4\%                           & NetBeans              & 4,740                                      & 314                                  & 6.6\%                           \\
Ambari           & 17,961                                     & 307                                  & 1.7\%                            & Hadoop           & 23,798                                     & 2,244                                & 9.4\%                            & NiFi                  & 4,427                                      & 221                                  & 5.0\%                           \\
Arrow            & 5,576                                      & 290                                  & 5.2\%                            & HBase            & 12,227                                     & 1,142                                & 9.3\%                            & OFBiz                 & 4,740                                      & 215                                  & 4.5\%                           \\
Axis2            & 4,554                                      & 425                                  & 9.3\%                            & Hive             & 14,306                                     & 509                                  & 3.6\%                            & Ozone                 & 2,471                                      & 194                                  & 7.9\%                           \\
Camel            & 5,758                                      & 807                                  & 14.0\%                           & Hudi             & 1,177                                      & 184                                  & 15.6\%                           & Qpid                  & 5,097                                      & 214                                  & 4.2\%                           \\
CloudStack       & 7,861                                      & 870                                  & 11.1\%                           & Ignite           & 7,242                                      & 582                                  & 8.0\%                            & Solr                  & 7,209                                      & 377                                  & 5.2\%                           \\
Cordova          & 8,780                                      & 813                                  & 9.3\%                            & Impala           & 6,238                                      & 1,431                                & 22.9\%                           & Spark                 & 15,265                                     & 1,773                                & 11.6\%                          \\
Drill            & 5,437                                      & 579                                  & 10.6\%                           & Jackrabbit-Oak   & 3,726                                      & 168                                  & 4.5\%                            & Thrift                & 3,058                                      & 183                                  & 6.0\%                           \\
Flink            & 10,296                                     & 2,049                                & 19.9\%                           & Kafka            & 6,808                                      & 617                                  & 9.1\%                            & Traffic Server        & 3,026                                      & 165                                  & 5.5\%                           \\
Geode            & 5,429                                      & 150                                  & 2.8\%                            & Lucene           & 4,201                                      & 201                                  & 4.8\%                            & Wicket                & 4,193                                      & 212                                  & 5.1\%                           \\
Groovy           & 6,918                                      & 352                                  & 5.1\%                            & Mesos            & 4,922                                      & 427                                  & 8.7\%                            & \textbf{All projects} & \textbf{223,355}                           & \textbf{18,440}                      & \textbf{8.3\%}                  \\ \hline
\end{tabular}
}
\end{table}

\subsection{Construction of Bug Fixing Evolution Features}
The fixing of software bugs is a complex and dynamically changing process. In this process, data related to dynamic changes in bugs can capture key features during the fixing process, which is crucial for predicting changes in bug priority. Specifically, we construct bug fixing evolution features from four perspectives: the project, the bug reporter, the bug comments, and the historical change items of the bug, in order to characterize the dynamic change features of bug fixing. 

\subsubsection{Features of the project to which the bug belongs}
Each bug belongs to a specific project, and as the project progresses, numerous bugs are resolved while new ones are reported. Consequently, the number of bugs that need to be addressed within the project is in a state of constant flux. This number directly reflects the current workload and pressure on the team, which can influence the setting of bug priorities. In this section, we will analyze two key features: the current count of unresolved bugs (with statuses ``OPEN,'' ``IN PROCESS,'' or ``REOPEN'') and their average priority. Understanding the project's current status is essential for predicting future changes in bug priorities. 
Supposing that the current project is denoted as $p$ and the time point is $t$, then ${Proj}_{p,t}I$ represents the current bug quantity feature of the project to which the bug belongs, which will be applied to the Phase I of bug priority change prediction. ${Proj}_{p,t}I$ is defined as below:

\begin{equation}
     {Proj}_{p,t}I = \left\{ N_{p,t},{ProjAvePri}_{p,t} \right\} ,
\end{equation}

\noindent where $N_{p,t}$ is the current number of bugs, and ${ProjAvePri}_{p,t}$ is the average priority of these unresolved bugs.
Let $d_{b}$ be the time interval from the bug being reported to the current time, ${NewN}_{d_{b}}$ be the number of bugs reported during that time interval, and ${CloseN}_{d_{b}}$ be the number of bugs resolved during that time interval. ${Proj}_{p,t}II$ represents the feature of the number of bugs within the bug fixing time interval in the project to which the bug belongs: 

\begin{equation}
{Proj}_{p,t}II = \left\{ {NewN}_{d_{b}},{CloseN}_{d_{b}} \right\} .
\end{equation}

\subsubsection{Features of a bug reporter}
There are four features from the perspective of a bug reporter, namely: the average and median priority of bugs previously reported by the bug reporter, the proportion of priority change directions and the average range of priority changes in the bugs previously reported by the bug reporter.
The experience of the reporter may affect their evaluation of bug priority. By calculating the priority mean and median of all bugs previously reported by the reporter, the general evaluation tendency of the reporter can be understood. 
Let $r$ be the current bug reporter, $t_c$ be the current time, and $B_{r,t_{c}} = \left\{ b_{1},~b_{2},b_{3},\ldots,b_{n} \right\}$ be the set of bugs reported by the bug reporter as of the current time, then the weight of $b_i$ is defined as:
\begin{equation}
w\left( b_{i} \right) = e^{- \frac{|t_{i} - t_{c}|}{T}} ,
\end{equation}
where $t_i$ denotes the reporting time of $b_i$, and $T$ denotes the time interval between the earliest bug reporting time in $B_{r,t_{c}}$ and the current time $t_c$. The weight $w\left( b_{i} \right)$ is set for each bug previously reported by the current bug reporter $r$ because the professional knowledge of the bug reporter will change over time \cite{SaSt2020}. If the time factor is not taken into account, there may be a problem of concept drift \cite{PaCaDi2021}. By assigning a weight $w\left( b_{i} \right)$ to each bug, bugs that are more distant in time have lower weights, thus better reflecting the reporter's current professional level. $Pri_i$ denotes the priority of $b_i$, hence the average priority of bugs reported by bug reporter $r$ before time $t_c$ is:
\begin{equation}
Rep{Ave}_{r,t_{c}} = {\sum\limits_{i}^{n}{\left( {w\left( b_{i} \right)*Pri}_{i} \right)/{\sum\limits_{i}^{n}{w\left( b_{i} \right)}}}} .
\end{equation}
The median priority of bugs reported by bug reporter $r$ before time $t_c$ is denoted as $Rep{Med}_{r,t_{c}}$. The median is taken when the bugs in $B_{r,t_{c}}$ are sorted by priority.
The overall priority change direction of the bug reporter before can reveal their habit of priority adjustment. 
Bugs that have undergone priority changes from set $B_{r,t_{c}}$ are selected to form the set $BPC_{r,t_{c}}$, the proportion of upward priority changes in $BPC_{r,t_{c}}$ is denoted as $RepUp_{r,t_{c}}$, and thus the proportion of downward changes is $1-RepUp_{r,t_{c}}$.

The range of priority change reflects the degree of change in the reporter's evaluation of bugs. A significant change in range may indicate that the reporter is not stable or uncertain enough when evaluating the priorities of bugs. 
Understanding the range of priority changes for reporters can better predict the trend of priority changes for new bugs. The average range of bug priority changes reported by bug reporter $r$ before time $t_c$ is:
\begin{equation}
Rep{AveRg}_{r,t_{c}} = {\sum\limits_{i}^{n}{\left( {w\left( b_{i} \right)*Rg}_{i} \right)/{\sum\limits_{i}^{n}{w\left( b_{i} \right)}}}} ,
\end{equation}
where $Rg_i$ represents the absolute value of the priority change range of $b_i$. Finally, the features of the bug reporter are summarized as follows:
\begin{equation}
{Rep}_{r,t_{c}}I = \left\{ Rep{Ave}_{r,t_{c}},Rep{Med}_{r,t_{c}} \right\}
\end{equation}
\begin{equation}
{Rep}_{r,t_{c}}II = \left\{ Rep{Ave}_{r,t_{c}},Rep{Med}_{r,t_{c}},Rep{Up}_{r,t_{c}},Rep{AveRg}_{r,t_{c}} \right\}
\end{equation}
${Rep}_{r,t_{c}}I$ and ${Rep}_{r,t_{c}}II$ will be applied to Phase I and II of bug priority change prediction, respectively. This means that the average and median of bug priority will be applied in both phases of bug priority change prediction, while the proportion and range of priority change direction will only be used in Phase II.

\subsubsection{Features of changes in bug comments}
A bug report contains a number of comments made by the bug participants, which record their communication during the bug fixing process and reflect their level of attention to the bug \cite{AbHe2023}. The features of changes in bug comments mainly include three aspects: the change rate in bug comment frequency, the change rate in bug commenter frequency, and the change rate in total bug comment length frequency. The number of bug comments will gradually increase with the progress of bug fixing activities, but the frequency of the increase in comment numbers will fluctuate. 
If the number of comments increases sharply during a certain period of time, it indicates that the attention to the bug has been increased, which may be due to the urgent need for handling the bug. For example, in the case of \href{https://issues.apache.org/jira/browse/AMQ-5430}{AMQ-5430}, this bug was initially reported on November 11, 2014. During the following nine days (from November 11 to November 19), only three additional comments were made. However, subsequently within just two days (November 25 and 26), five more comments were added, and the priority of the bug was eventually modified on November 26. The change in frequency of the increase in the number of comments reflects the differences in frequency changes between different time periods, and the other two features (the change rate in bug commenter frequency, and the change rate in total bug comment length frequency) are also defined based on the similar logic.

\begin{figure*}
    \centering{\includegraphics[width=5.5in]{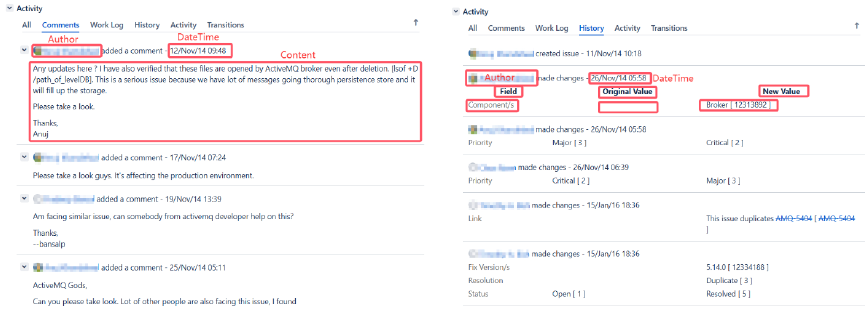}}
    \caption{Screenshot of some comments and history items for bug \href{https://issues.apache.org/jira/browse/AMQ-5430}{AMQ-5430} of the ActiveMQ project. }
    \label{fig:comment_history}
\end{figure*}

\begin{figure*}
    \centering{\includegraphics[width=5.0in]{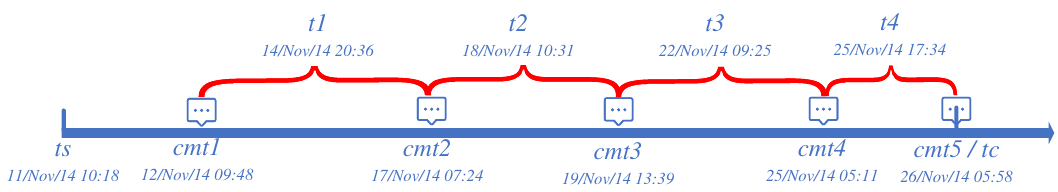}}
    \caption{The comments in the timeline for bug \href{https://issues.apache.org/jira/browse/AMQ-5430}{AMQ-5430} of the ActiveMQ project. }
    \label{fig:diff}
\end{figure*}

We take \href{https://issues.apache.org/jira/browse/AMQ-5430}{AMQ-5430} as an example to illustrate the types of comment-related data we collected. Fig. \ref{fig:comment_history} presents a screenshot of the bug’s comments (left) and its historical item changes (right). For each bug, we collected all associated comments and extracted the data items highlighted in the red boxes in Fig. \ref{fig:comment_history}, such as the author, datetime, and content of each comment. Subsequently, Fig.~\ref{fig:diff} provides an example to demonstrate how we calculated the variation rate of bug comment frequency. It demonstrates the timeline of comments posted from the reporting time of bug \href{https://issues.apache.org/jira/browse/AMQ-5430}{AMQ-5430} until its first priority change. Time points $ts$ and $tc$ denote the time the bug was reported and the current time, respectively. During the time interval, a total of five comments were posted, and Fig. \ref{fig:diff} indicates the exact time points of these comments. First, calculate the time midpoint between the first two comments, $cmt1$ and $cmt2$, denoted as $t1$, using $t1$ as the dividing point. Before $t1$, there is one comment, with the interval between $t1$ and the reporting time being 3.43 days (i.e., from 11/Nov/14 10:18 to 14/Nov/14 20:36). Between $t1$ and $tc$ there are 4 comments, spanning an interval of 11.42 days (i.e., from 14/Nov/14 20:36 to 26/Nov/14 05:58). The comment frequency before $t1$ is thus 1/3.43 = 0.29, while the comment frequency after $t1$ is 4/11.42 = 0.35. The change in comment frequency after $t1$ relative to the prior frequency is (0.35-0.29)/0.29 = 0.21, denoted as $NumDiff1$. Similarly, $NumDiff2$, $NumDiff3$ and $NumDiff4$ are calculated as 0.34, 0.88, and 5.57, respectively. The change rate in bug comment frequency, denoted as $maxNumDiff$ is the maximum absolute value among $NumDiff1$ to $NumDiff4$. In a similar manner, the change rate in bug commenter frequency and the change rate in total bug comment length frequency can be calculated, but the calculation unit has been changed from the number of comments to the number of commenters and the total length of comments. These two features are respectively referred to as $maxPersDiff$ and $maxLenDiff$. 

Thus, at $t_c$, the features in comments changes of $b$ are:
\begin{equation}
    {Cmt}_{b,t_{c}}II = \left\{ maxNumDiff,maxPersDiff,maxLenDiff\right\}
\end{equation}
The comment change feature is only applicable to Phase II of bug priority change prediction, as in Phase I, there is no comment data when bugs are just reported. 

\subsubsection{Features of changes in bug history items}
During the bug fixing process, the fields of a bug may be modified, and the records of these modifications reflect the historical change information of the bug. As shown in Fig.~\ref{fig:comment_history}, the right side presents a screenshot of the historical change records for \href{https://issues.apache.org/jira/browse/AMQ-5430}{AMQ-5430}. We collected all history change items for each bug, and for each item, we extracted the data highlighted in the red boxes in Fig.~\ref{fig:comment_history}, including the author of the change, the datetime, the modified field, the original value of the field, and its new value. It is worth noting that we gathered all field changes, which primarily include: \textit{Summary}, \textit{Description}, \textit{Status}, \textit{Resolution}, \textit{Type}, \textit{Assignee}, \textit{Environment}, \textit{Labels}, \textit{Affects Version/s}, \textit{Fix Version/s}, \textit{Component/s}, \textit{Attachment/s}, \textit{Link}, and \textit{Remote Link}. After collecting these history change items, we further extracted features related to the bug's historical changes. The features of bug history change items include: the change rate in bug history item frequency, the change rate in bug history participant frequency, and the change rate in bug history field change frequency. 

Let $t_s$ be the reporting time of bug $b$, $t_c$ be the current time, and ${Hists}_{b,t_{c}} = \left\{ h_{1},~h_{2},h_{3},\ldots,h_{n} \right\}$ be the historical change item set of $b$ by deadline $t_c$, the three features of bug historical change items are the same as the calculation process of bug comment change features in Fig. \ref{fig:diff}. Only the calculation unit needs to be replaced with the number of bug history change items, the number of bug history change participants, and the number of bug history change fields. Therefore, the calculation process of bug historical change features will not be repeated here. Thus, when $b$ is at $t_c$, its historical change features are:
\begin{equation}
    {Hist}_{b,t_{c}}II = \left\{ maxItemDiff,maxAuthorDiff,maxFieldDiff\right\} ,
\end{equation}
where $maxItemDiff$ denotes the maximum change rate in bug history change item frequency, $maxAuthorDiff$ denotes the maximum change rate in bug history participant frequency, and $maxFieldDiff$ denotes the maximum change rate in bug history field change frequency. Similar to the feature of comment changes, the bug history change item feature is only applicable to Phase II of priority change prediction, as there are no historical change data when the bug is just reported.

\subsubsection{Bug Fixing Evolution Features}
By combining the features collected from four perspectives, i.e., project, bug reporter, bug comments, and bug history changes, we can obtain bug fixing evolution features. The evolution features of bug fixing for $b$ at $t_c$ can be expressed as:
\begin{equation}
{BFE}_{b,t_{c}}I = concat\left( {{Proj}_{p,t}I,Rep}_{r,t_{c}}I \right) ,
\end{equation}
\begin{equation}
{BFE}_{b,t_{c}}II = concat\left( {{Proj}_{p,t}II,Rep}_{r,t_{c}}II,{Cmt}_{b,t_{c}}II,{Hist}_{b,t_{c}}II \right) ,
\end{equation}
where $concat()$ is a connection operation that combines various features together. ${BFE}_{b,t_{c}}I$ will be applied to Phase I of bug priority change prediction, and ${BFE}_{b,t_{c}}II$ will be applied to Phase II.

\subsection{Basic Features and Textual Features of Bugs}

\begin{table}[]
\centering
\caption{Basic features of a bug}
\label{table:basicf}
\scalebox{0.8}{
\begin{tabular}{p{0.12\columnwidth} p{0.14\columnwidth} p{0.07\columnwidth} p{0.8\columnwidth} }
\hline
\textbf{Category}                     & \textbf{Feature}    & \textbf{Phase} & \textbf{Description}                                                                                                                                                                                                                                                                \\ \hline
Project                      & ProjId     & I, II & The project to which the bug belongs.                                                                                                                                                                                                                                       \\ \hline
\multirow{2}{*}{Summary}     & SmyLen     & I, II & The length of the summary content for the bug, in bytes.                                                                                                                                                                                                                    \\
                             & SmyChanged & II    & Has the summary of this bug been modified since the reporting time till now? This is a Boolean feature.                                                                                                                                                                             \\ \hline
\multirow{2}{*}{Description} & DesLen     & I, II & The length of the description content for the bug, in bytes.                                                                                                                                                                                                                \\
                             & DesChanged & II    & Has the description of the bug been modified since the reporting time till now?                                                                                                                                                                                                    \\ \hline
\multirow{6}{*}{Related Bug} & RelNum     & I, II & The number of related bugs. In this study, the related bugs of a bug are defined as follows: bugs with the same component, version, fix version, label, attachment, and bugs in the Link field of the bug. 
\\
                             & RelPCNum   & I, II & The number of changes in priority of related bugs.                                                                                                                                                                                                                         \\
                             & RelAve     & I, II & Mean priority of related bugs.                                                                                                                                                                                                                                             \\
                             & RelMed     & I, II & Median priority of related bugs.                                                                                                                                                                                                                                           \\
                             & RelUp      & II    & Proportion of upward priority change of related bugs.                                                                                                                                                                                                                      \\
                             & RelRange   & II    & Average range of changes in priority of related bugs.                                                                                                                                                                                                                      \\ \hline
\multirow{3}{*}{Reporter}    & RepId      & I, II & The identity of the bug reporter.                                                                                                                                                                                                                                                                           \\
                             & RepBugNum  & I, II & The bug reporter has reported the number of bugs.                                                                                                                                                                                                                          \\
\multirow{5}{*}{Priority}    & CurPri     & I, II & Bug current priority.                                                                                                                                                                                                                                                      \\
                             & RepPCNum   & I, II & The bug reporter has reported the number of bugs with priority changes.                                                                                                                                                                                                     \\ \hline
                             & Changed    & II    & Has there been a priority change for this bug since the reporting time till now?                                                                                                                                                                                             \\
                             & PCNum      & II    & The number of priority changes underwent by this bug from being reported to the current one.                                                                                                                                                                               \\
                             & PCDir      & II    & The direction of the current priority of the bug compared to the initial priority.                                                                                                                                                                                         \\
                             & PCRange    & II    & The change range in the current priority of the bug compared to the initial priority (absolute value).                                                                                                                                                                     \\ \hline
\end{tabular}}
\end{table}

In addition to utilizing bug fixing evolution features, we also selected several basic attributes as basic features of a bug. These attributes include project ID, current priority, summary, description, and other bug information related to the bug. The selection of certain features is based on the work of Tian \textit{et al} \cite{TiLoXiSu2015}. The basic features help the model understand the complexity and level of detail in the bug descriptions \cite{YaZhLe2014}. Table \ref{table:basicf} lists the basic features of the bugs selected in this study. In this table, each feature is classified according to its category, and specific feature names are listed. It also specifies at which phase of the bug priority change prediction process these functions should be applied. Some features are only applied in Phase II (e.g., SmyChanged) because these features can only be collected during the bug-fixing process and are not available when the bug is just reported. Finally, a description of each feature is provided. 

The Summary and Comment fields of the bug report contain rich textual information that provides specific contextual descriptions, which are crucial for assessing the scope and priority of the bug's impact. The Summary field is equivalent to the title of a bug report, usually summarizing the core problem of the bug in a short sentence; therefore, this field contains important bug information. The Comment field refers to the comment content of the bug, which reflects the communication and decision-making of the participants in the process of fixing the bug \cite{GaWaHeZhZhLy2015}. We used a pre-trained RoBERTa model to extract text features with a default length of 768. We chose RoBERTa to extract text features because the BERT series of models can generate context-sensitive word embeddings. Many studies have also demonstrated that for tasks requiring a deep semantic understanding of software engineering artifacts - such as bug prediction \cite{GoDaCo2023}, bug assignment \cite{WajiXuTiJiHu2024}, and bug classification \cite{Ku2024} - BERT-family models substantially outperform simpler alternatives like TF-IDF and Word2Vec. This ability to understand context is crucial for accurately assessing the priority of bug reports. RoBERTa is a direct improvement to BERT, and previous studies have shown that RoBERTa has achieved good results in bug priority prediction (e.g., \cite{IzAkHe2022}). 


\subsection{Class Imbalance Handling Strategy}
There is a serious class imbalance in the priority of bug reports, which can lead to the neglect of minority class features and poor generalizability for new minority class samples. In view of this, we will explore how to effectively solve the problem of class imbalance from the perspective of data processing and algorithm optimization at each phase, in order to improve the performance of minority class prediction. Specifically, we first verify and illustrate the motivation and rationale of the selected category imbalance handling strategy through a series of pilot experiments. Then we provide a detailed description of the specific strategies adopted in Phase I and Phase II and their implementation details.

\subsubsection{Pilot Experiments}
\label{sec:pilot}

To motivate the imbalance handling strategies adopted in our approach, we conducted a series of pilot experiments prior to the main study. The goal of these experiments was not to exhaustively optimize hyper-parameters, but rather to provide empirical evidence for the design choices in Phase~I and Phase~II. All pilot experiments were implemented with the \texttt{imblearn}, \texttt{scikit-learn}, and \texttt{xgboost} libraries, using their default parameter configurations unless otherwise specified.

\paragraph{Phase I}
We first compared various sampling methods for the binary classification task in Phase~I. Table~\ref{tab:pilot-sampling-phase1} summarizes the results. It can be observed that K-means undersampling achieved the best F1-score among all candidates. Unlike random or neighbor-based undersampling (e.g., NearMiss, Tomek Links, ENN), K-means can preserve the diversity of majority class samples by selecting cluster representatives, which explains its superior performance. Therefore, we adopted K-means undersampling in Phase~I.

\begin{table}[h]
\centering
\caption{Pilot results of different sampling methods in Phase I.}
\label{tab:pilot-sampling-phase1}
\scalebox{0.8}{
\begin{tabular}{p{0.2\columnwidth} p{0.3\columnwidth} p{0.5\columnwidth} p{0.1\columnwidth} }
\hline
\textbf{Sampling Type} & \textbf{Method} & \textbf{Parameters} & \textbf{F1-score} \\
\hline
No sampling & None & -- & 0.07 \\
\hline
Undersampling & Random undersampling & -- & 0.59 \\
              & NearMiss & version=1, n\_neighbors=3, sampling\_strategy=auto & 0.63 \\
              & Tomek Links & sampling\_strategy=auto & 0.08 \\
              & ENN & n\_neighbors=3, kind\_sel=all, sampling\_strategy=auto & 0.07 \\
              & K-means & init=k-means++, n\_init=10, max\_iter=300, tol=1e-4, algorithm=lloyd & \textbf{0.66} \\
\hline
Oversampling & Random oversampling & -- & 0.62 \\
             & SMOTE & sampling\_strategy=auto, k\_neighbors=5 & 0.61 \\
\hline
\end{tabular}
}
\end{table}

Next, we compared several machine learning algorithms, as shown in Table~\ref{tab:pilot-models-phase1}. Random Forest, KNN, SVM, and XGBoost achieved relatively high F1-scores. Although Decision Tree also performed reasonably well, its predictive behavior is highly correlated with Random Forest, and thus it was excluded to avoid redundancy. Logistic Regression and Gaussian Naive Bayes showed weaker performance, and were not considered further. Therefore, we retained four classifiers: Random Forest, KNN, SVM, and XGBoost.

\begin{table}[h]
\centering
\caption{Pilot results of different classifiers in Phase I.}
\label{tab:pilot-models-phase1}
\scalebox{0.8}{
\begin{tabular}{p{0.2\columnwidth} p{0.7\columnwidth} p{0.1\columnwidth} }
\hline
\textbf{Model} & \textbf{Parameters} & \textbf{F1-score} \\
\hline
Logistic Regression & penalty=l2, C=1.0, solver=lbfgs, max\_iter=100 & 0.63 \\
Decision Tree & criterion=gini, splitter=best, max\_depth=8 & 0.65 \\
Random Forest & n\_estimators=100, criterion=gini, max\_features=sqrt & \textbf{0.73} \\
Gaussian Naive Bayes & priors=None, var\_smoothing=1e-9 & 0.59 \\
KNN & n\_neighbors=5, weights=uniform, metric=minkowski & 0.66 \\
SVM & C=1.0, kernel=rbf, gamma=scale & 0.68 \\
XGBoost & max\_depth=6, learning\_rate=0.3, n\_estimators=100, objective=binary:logistic & 0.72 \\
\hline
\end{tabular}
}
\end{table}

Finally, we evaluated ensemble strategies using the four selected classifiers. Hard voting achieved an F1-score of 0.74, while soft voting (with equal weights) yielded 0.77. As soft voting better balances model uncertainties and provided higher performance, we adopted the soft voting strategy in Phase~I.

\paragraph{Phase II}
Following a similar procedure, we performed pilot experiments in Phase~II to compare oversampling, undersampling, and mixed sampling strategies. The \texttt{k\_neighbors} of SMOTE uses the default value, which is 5. The results are summarized in Table~\ref{tab:pilot-sampling-phase2}. It can be seen that mixed sampling yielded the best results. Two clarifications are necessary: (1) the goal of Phase~II pilots was to compare different types of sampling (rather than identifying the single best technique), and thus we used representative baselines (random undersampling and SMOTE oversampling); (2) given the multi-class nature of Phase~II, trivial and minor classes were oversampled, while other majority classes were undersampled. We performed a simple grid search over sampling ratios to balance performance and computational cost, and selected the relatively best configuration.

\begin{table}[h]
\centering
\caption{Pilot results of different sampling methods in Phase II.}
\label{tab:pilot-sampling-phase2}
\scalebox{0.8}{
\begin{tabular}{lll}
\hline
\textbf{Method} & \textbf{Strategy} & \textbf{F1-weighted} \\
\hline
No sampling & -- & 0.45 \\
SMOTE oversampling & Trivial +100\%, Minor +50\% & 0.49 \\
Random undersampling & Blocker/Critical/Major -10\% & 0.52 \\
                     & Blocker/Critical/Major -20\% & 0.58 \\
                     & Blocker/Critical/Major -30\% & 0.55 \\
Mixed sampling & Trivial +100\%, Minor +50\%; Blocker/Critical/Major -20\% & \textbf{0.62} \\
\hline
\end{tabular}
}
\end{table}

\subsubsection{Phase I}
As shown in Table \ref{table:proportion}, we collected bug reports for 32 non-trivial Apache projects from JIRA, totaling 223,355 bug reports. Among them, only \textbf{18,440} bugs (\textbf{8.3\%}) experienced priority changes during their lifecycle, whereas the remaining \textbf{204,915} bugs (\textbf{91.7\%}) did not. This results in a highly imbalanced distribution between the positive and negative classes, with a ratio of approximately \textbf{1:11}.
To address class imbalance, we first performed undersampling using the K-means algorithm at the data level \cite{LiTsHuJh2017}. We clustered the majority class samples into 18,440 groups, equaling the number of minority class samples. We then selected these 18,440 centroids as representative samples to replace the original majority class samples, as it captures the main features of each cluster sample. 
At the algorithm level, we used model ensembling to construct a bug priority change prediction model. Specifically, we employed four machine learning algorithms—RF, KNN, SVM, and XGBoost—and combined them using soft voting. By ensembling them, we leveraged their complementary strengths to enhance the overall generalizability \cite{DoYuCaShMa2020}. Soft voting involves each base model providing a probability distribution, allowing the final prediction to be a weighted average, thus making predictions more robust. In Phase I, we chose to use K-means undersampling at the data level and ensemble learning at the algorithm level \cite{Ma1967, DoYuCaShMa2020}. This decision was made based on pilot experiments that showed that the two methods worked best when used in combination. Specifically, K-means undersampling helps balance the distribution of class in the dataset, while ensemble learning enhances the stability and predictive power of the model. Therefore, considering these advantages, we adopted this combination strategy in Phase I.

\subsubsection{Phase II}\label{sec:imblance_2}
We counted 18,440 bugs that experienced priority changes, resulting in 21,215 priority change records, as some bugs had multiple changes. Table \ref{table:PCDistribution} shows the distribution of these changes. The left side represents the initial priority, and the top represents the changed priority. Darker cells indicate a higher number of bug samples in that class. Although the priority distribution after the change is generally more uniform (except for the small number of Trivial priorities), the initial priority distribution is imbalanced. Specifically, the initial priority ``Major'' accounts for the majority of changes (\textbf{12,746 records, 60\% of all changes}), while the number of changes originating from ``Trivial'' is extremely small (\textbf{only 369 records, less than 2\%}). Li et al. found that the bug priority change is actually significantly influenced by the initial priority \cite{LiCaYuLiMoLi2024}. Therefore, when predicting priority changes during the bug fixing phase, we will focus on data processing and algorithm optimization to mitigate the adverse effects of class imbalance.

\begin{table}[]
\centering
\caption{Distribution of changes with different bug priorities.}
\label{table:PCDistribution}
\scalebox{0.80}{
\begin{tabular}{lrrrrrr}
\hline
                  & \multicolumn{1}{c}{\textbf{Blocker}} & \multicolumn{1}{c}{\textbf{Critical}} & \multicolumn{1}{c}{\textbf{Major}} & \multicolumn{1}{c}{\textbf{Minor}} & \multicolumn{1}{c}{\textbf{Trivial}} & \multicolumn{1}{c}{\textbf{All}} \\
\textbf{Blocker}  & 0                                    & \cellcolor[HTML]{ADADAD}1,019                                  & \cellcolor[HTML]{ADADAD}1,480                               & \cellcolor[HTML]{D9D9D9}298                                & \cellcolor[HTML]{EFEFEF}29                                   & 2,826                             \\
\textbf{Critical} & \cellcolor[HTML]{ADADAD}1,419                                 & 0                                     & \cellcolor[HTML]{ADADAD}1,789                               & \cellcolor[HTML]{D9D9D9}370                                & \cellcolor[HTML]{EFEFEF}19                                   & 3,597                             \\ 
\textbf{Major}    & \cellcolor[HTML]{808080}3,755                                 & \cellcolor[HTML]{808080}4,266                                  & 0                                  & \cellcolor[HTML]{808080}4,383                               & \cellcolor[HTML]{D9D9D9}342                                  & 12,746                            \\
\textbf{Minor}    & \cellcolor[HTML]{D9D9D9}209                                  & \cellcolor[HTML]{D9D9D9}312                                   & \cellcolor[HTML]{ADADAD}988                               & 0                                  & 168                                  & 1,677                             \\ 
\textbf{Trivial}  & \cellcolor[HTML]{EFEFEF}15                                   & \cellcolor[HTML]{EFEFEF}15                                    & \cellcolor[HTML]{EFEFEF}72                                 & \cellcolor[HTML]{D9D9D9}267                                & 0                                    & 369                              \\
\textbf{All}      & 5,398                                 & 5,612                                  & 4,329                               & 5,318                               & 558                                  & 21,215                            \\ \hline
\end{tabular}%
}
\end{table}

At the data level, we employed the conditional mixed sampling method to address class imbalance. This method ensures that the probability distribution of each priority change class remains consistent with the original data when using the SMOTE oversampling and random undersampling techniques. The procedure of conditional mixed sampling is as follows:

\begin{enumerate}
    \item \textbf{Data Partitioning by Initial Priority} \\
    The entire training dataset is first partitioned into five subsets according to the initial priority levels. All subsequent sampling operations are performed independently within each subset.

    \item \textbf{Determining Sampling Rates for Each Class} \\
    Based on our pilot experiment, we conducted a grid search to identify the relatively optimal over-sampling and under-sampling rates for each initial priority class. These rates are empirically chosen to enhance the signal of underrepresented categories without significantly distorting the overall data distribution. Detailed parameters are provided in Table~\ref{table:sampling_ratios}.

    \item \textbf{Iterative Application of Sampling Techniques} \\
    Each subset is processed according to its designated sampling class:
    \begin{itemize}
        \item For classes marked for over-sampling (e.g., ``Minor'' and ``Trivial''), the SMOTE (Synthetic Minority Over-sampling Technique) algorithm is applied. For each original sample, SMOTE generates synthetic samples by interpolating between the sample and its $k$ nearest neighbors within the same subset. The number of generated samples is determined by the pre-defined over-sampling rate and strictly follows the conditional constraints described in Step 4.
        \item For classes marked for under-sampling (e.g., ``Major''), random under-sampling is performed by removing a specified percentage of samples to reduce the dominance of the category.
    \end{itemize}

    \item \textbf{Enforcing Conditional Change Distribution Constraints} \\
    This step captures the ``conditional'' characteristic of our approach and ensures the authenticity of the generated data. Specifically, when SMOTE generates new synthetic samples for a given initial priority (e.g., a new ``Minor'' bug), the target priority (i.e., the updated priority) is not chosen randomly. Instead, it is sampled according to the empirical change probability distribution observed for that initial priority class.

    \item \textbf{Reassembling the Dataset} \\
    Finally, the synthetic samples from the over-sampled subsets, the remaining samples from the under-sampled subsets, and the samples from any unsampled subsets are combined to form the final rebalanced training dataset. This dataset is then used to train the second-stage prediction model.
\end{enumerate}

\begin{table}[htbp]
\centering
\caption{Sampling Ratios for Each Initial Priority Class}
\label{table:sampling_ratios}
\scalebox{0.8}{
\begin{tabular}{p{0.2\columnwidth} p{0.2\columnwidth} p{0.75\columnwidth} }
\hline
\textbf{Parameter} & \textbf{Value Used} & \textbf{Description \& Rationale} \\
\hline
Initial `Blocker' & +15\% Oversampling & Boosts the signal for this critical but less frequent starting class. \\
Initial `Critical' & +15\% Oversampling & Similar to `Blocker', enhances representation. \\
Initial `Major' & -10\% Undersampling & Reduces the dominance of the most over-represented starting class. \\
Initial `Minor' & +65\% Oversampling & Significantly increases the number of samples for this under-represented class. \\
Initial `Trivial' & +80\% Oversampling & Aggressively oversamples the rarest starting class to make its patterns learnable. \\
\hline
\end{tabular}
}
\end{table}

Similar to Phase I, Phase II also uses conditional mixed sampling based on the results of pilot experiments. It is important to note that the term ``conditional'' refers to our sampling strategy being constrained to preserve the original change distribution for each initial priority category. For example, in the original dataset, suppose there are 1,000 bugs with the initial priority ``Blocker'', of which 200 (20\%) are eventually reassigned to ``Critical''. Our conditional sampling ensures that this 20\% proportion is retained. Concretely, if we oversample the ``Blocker'' category by generating 500 synthetic samples, exactly 100 of them (20\%) will also represent the change from ``Blocker'' to ``Critical''. The same principle applies to all possible priority transitions, thereby preventing the sampling process from distorting the original change structure. The advantage of conditional mixed sampling lies in its ability to balance both quantity and data authenticity. On the one hand, its ``conditional'' design ensures that the distribution structure of each priority change is preserved during the resampling process, thereby avoiding the introduction of unreasonable samples that violate the original distribution. On the other hand, its ``mixed'' mechanism allows for flexible setting of sampling intensity based on class characteristics: moderate expansion of the minority class while reasonable compression of the majority class. This differentiated strategy effectively avoids the problems of minority classes being neglected and majority classes being overly dominant, and is more targeted and robust than relying solely on a single sampling method.

At the algorithm level, we adopted cost-sensitive learning to adjust the importance of different priority classes. Despite conditional mixed sampling, the dataset remains imbalanced due to the few Trivial bugs. To address this, we implemented cost-sensitive learning with a weighted loss function. This assigns higher weights to prediction errors for minority classes, helping the model better learn their features and improve predictive performance. Specifically, we set the weight of each priority class to the reciprocal of its proportion of the total number of samples in the dataset.

        
        

After calculating the weights of each priority class, we add these weights to the loss function to calculate the final loss value, which helps reduce the prediction error of the model. In Phase II, since the priority change is considered a multi-classification problem, we choose the cross entropy loss function as the loss function to measure the performance of the model, and its specific formula is defined in Formula (\ref{eq:loss}): 
\begin{equation}\label{eq:loss}
    L = - \frac{1}{batch\_ size}{\sum\limits_{j = 1}^{batch\_ size}{\sum\limits_{i = 1}^{n}{w_{i}\left\lbrack y_{j,i}log{\hat{y}}_{j,i} - \left( 1 - y_{j,i} \right)log\left( 1 - {\hat{y}}_{j,i} \right) \right\rbrack}}} ,
\end{equation}
where $batch\_size$ is the number of samples, $n$ denotes the total number of priority classes, $w_i$ is the weight of the i-th priority class, $y_{j,i}$ is the true label of the j-th sample belonging to category $i$ (with a value of 0 or 1), and ${\hat{y}}_{j,i}$ refers to the model's prediction probability that the jth sample belongs to category $i$.

\subsection{Construction of Bug Priority Change Prediction Model}
We built distinct prediction models for Phase I and Phase II separately. Firstly, we integrated the evolution features of bug fixing, basic features of bugs, and text features of bugs to form a dataset for predicting bug priority changes. The resulting feature vectors include ${PCVec}_{b,t_{c}}I$ and ${PCVec}_{b,t_{c}}II$. The former is used to construct a prediction model for Phase I, while the latter is applied to construct a prediction model for Phase II. Specifically, as shown in Formulas (\ref{eq:PCV1}) and (\ref{eq:PCV2}), where ${BFED}_{b,t_{c}}I$ denotes the bug fxing evolution features of $b$ at $t_c$, $Basic{Vec}_{b,t_{c}}I$ covers the basic bug features of Phase I in Table \ref{table:basicf}, $TextVec_{b,t_{c}}I$ is the bug report text features extracted through RoBERTa, and for ${PCVec}_{b,t_{c}}II$, its constituent elements follow the same logic. For clarity, we emphasize how the evolution features are utilized in each prediction phase. In Phase I, the evolution features are directly concatenated with static and textual features to construct the final feature vector (${PCVec}_{b,t_{c}}I$), which is then fed into machine learning models to determine whether a priority change will occur. In Phase II, richer evolution features are incorporated into the feature vector to characterize the evolution patterns of priority change. The feature vector (${PCVec}_{b,t_{c}}II$) is subsequently input into a neural network model to predict the specific target priority. By combining evolution features with static and textual features, the model is able to more accurately distinguish among potential target priority categories.
\begin{equation}\label{eq:PCV1}
{PCVec}_{b,t_{c}}I = concat\left( {BFE}_{b,t_{c}}I,{Basic{Vec}_{b,t_{c}}I,TextVec}_{b,t_{c}}I \right)
\end{equation}
\begin{equation}\label{eq:PCV2}
    {PCVec}_{b,t_{c}}II = concat\left( {BFE}_{b,t_{c}}II,{Basic{Vec}_{b,t_{c}}II,TextVec}_{b,t_{c}}II \right)
\end{equation}
In Phase I, our goal is to develop a prediction model using ML to determine whether the priority of a bug will change. We will employ four ML algorithms: RF, KNN, SVM, and XGBoost. Model ensembling is achieved through a soft voting strategy, where each classifier outputs a prediction probability, which is then weighted. The weighted probabilities are summed, and the class with the highest total probability is chosen as the final prediction result. We use ${PCVec}_{b,t_{c}}I$ as input data to train the models mentioned above, and determine the optimal weight configuration for each classifier in the soft voting ensemble through grid search technology. We selected four machine learning models (Random Forest, KNN, SVM, and XGBoost) in Phase I based on the following considerations:
1) Sufficient data: Our dataset is large enough to provide sufficient training samples for these models to fully utilize their learning capabilities.
2) High data quality: There are fewer null values in the data, which is conducive to the application of machine learning models, because many high-performance models cannot directly handle missing values.
3) Moderate task complexity: Given that the prediction task in Phase I is relatively simple, theoretically, models with simpler structures can be adopted. However, to ensure prediction performance, we still selected models that perform well in similar tasks.
The reason for selecting these four specific models is that they perform relatively better in similar application scenarios, such as bug priority prediction \cite{AlBa2013, ChLiGuXuZh2020, MaDaHaPa2021}, especially Random Forest and XGBoost, which are specially selected for their advantages in processing large-scale datasets. In contrast, other models such as Naive Bayes were not considered after being evaluated in pilot experiments due to their poor performance in this research scenario.

In Phase II, we developed an artificial neural network to predict changes in bug priority. The neural network architecture includes an input layer, three hidden layers, and an output layer. The hidden layers capture complex nonlinear relationships between input features, enhancing model expressiveness. Both the input and hidden layers use ReLU activation functions to facilitate nonlinear mappings and mitigate the vanishing gradient problem. The output layer uses the Softmax function to convert the model's output into a probability distribution, providing prediction probabilities for each class. During training, we applied a cost-sensitive learning method defined in Formula (\ref{eq:loss}) as the loss function. Through backpropagation, the model iteratively adjusts its parameters to minimize the loss function, thereby improving the accuracy and robustness of its predictions regarding bug priority changes. We chose to use deep learning technology in Phase II mainly based on the following considerations: The task complexity of Phase II is relatively high, requiring the model to capture the specific relationships in the data to achieve accurate predictions. In contrast, deep learning models, with their powerful representation learning capabilities, can extract high-level abstract features from a large number of statistical features, thereby enhancing the model's generalization ability and prediction performance.

\subsection{Baseline Method}
As bug priority change prediction is a novel task, distinct from traditional bug priority prediction, existing methods are not directly applicable. To rigorously evaluate our proposed two-phase model and provide a meaningful performance baseline, we designed and implemented a heuristic baseline based on historical frequencies. It is crucial to clarify that this baseline is not a simple random sampler; it is a data-driven, probabilistic model designed to replicate the historical statistical patterns present in the training data. It does not make uniform random guesses. Instead, its predictions are weighted by the actual, observed probabilities learned from the dataset. This baseline is intended to simulate a simple yet rational decision-making process that relies solely on statistical patterns in the data, without leveraging the complex, dynamic evolutionary features that are central to our method.

We chose this heuristic baseline over adapting models from traditional priority prediction due to two fundamental challenges. First, feature space mismatch: existing static models are ill-equipped to handle the dynamic, event-driven nature of the priority change prediction task without substantial re-engineering. Second, evaluation ambiguity: it is unclear how to score a ``correct'' priority prediction from an adapted model when no actual priority change has occurred. Our probabilistic heuristic avoids these issues, providing a clear and interpretable ``lower bound'' that directly measures the performance gain attributable to our models' ability to learn from dynamic evolutionary features of bugs, rather than merely replicating historical frequencies of bugs.

\textbf{Baseline for Phase I:} Acknowledging that the propensity for change varies significantly across different initial priorities \cite{LiCaYuLiMoLi2024}, we designed a probability predictor for the binary classification task in Phase~I. It operates as follows:
\begin{enumerate}
    \item \textbf{Learning:} First, the historical probability of a change is calculated for each initial priority level $p_i$ from the entire training dataset, yielding a mapping $P(\text{Change} | \text{Initial\_Priority} = p_i)$.
    \item \textbf{Prediction:} For each bug in the test set, a random number $r$ is drawn from a uniform distribution $[0, 1)$. If $r$ is less than the pre-calculated change probability corresponding to the bug's initial priority, the baseline predicts a ``Change'' (1); otherwise, it predicts ``No Change'' (0).
\end{enumerate}

\textbf{Baseline for Phase II:} For the multi-class prediction task in Phase~II, we extended the heuristic logic to a predictor based on the historical change probability matrix. It is constructed and operates as follows:
\begin{enumerate}
    \item \textbf{Learning:} A change probability matrix is learned from the training data. This matrix defines the probability distribution of changing from an initial priority $p_i$ to any target priority $p_j$, denoted as $P(\text{Target} = p_j | \text{Current} = p_i)$.
    \item \textbf{Prediction:} For each bug in the test set, the predictor identifies the probability distribution corresponding to its initial priority $p_i$ from the change matrix. It then performs a random draw from the set of all possible target priorities according to this distribution, with the drawn sample serving as the prediction.
\end{enumerate}

This baseline represents the performance ceiling achievable by merely replicating historical statistical patterns. It provides a fair ``lower bound'' for evaluating our model's ability to learn complex patterns from deep features.

\section{Evaluation}\label{sec_evaluation}
\subsection{Research Questions}\label{sec:rq}
In order to comprehensively evaluate the performance and practical application value of our proposed bug priority change prediction model, we formulated the following three research questions (RQs) to evaluate our prediction method and further refine the third RQ as follows.
\begin{itemize}
\setlength{\itemsep}{0pt}
\setlength{\parsep}{0pt}
\item [\textbf{RQ1:}] \textbf{To what extent do our proposed bug fixing evolution features and class imbalance handling strategy contribute to the performance of bug priority change prediction?} This RQ aims to verify the specific contributions of the aforementioned methods to improving model performance through a series of ablation experiments, thereby confirming their feasibility and potential advantages in practical applications. 

\item [\textbf{RQ2:}] \textbf{How does our proposed class imbalance handling strategy compare to other widely-used sampling and algorithmic methods in predicting bug priority changes?} This RQ is dedicated to evaluating and comparing the differences and advantages and disadvantages between our class imbalance handling strategy and other common handling techniques. Through detailed comparative analysis, we will be able to determine whether the proposed class imbalance handling strategy can demonstrate clear superiority in relevant evaluation metrics.

\item [\textbf{RQ3:}] \textbf{How effective is our proposed prediction method when applied in a cross-project context and across different priority levels?} To comprehensively evaluate the practical application performance of our proposed bug priority change prediction method, we will address this RQ from two perspectives. Firstly, we will test the generalizability and robustness of the predictive model through cross project testing; Secondly, we evaluate the performance of the prediction model for different priorities under different priority conditions.
\end{itemize}

\subsection{Experimental Setup}
To train the bug priority change prediction model for Phase I, we adopted a model ensemble approach using four ML algorithms: RF, KNN, SVM, and XGBoost. The parameter configurations for these classifiers are detailed in Table \ref{table:parameter}. Using grid search with a total weight of 10, where the weight search range for each classifier was 1 to 7 and the step size was 1, the optimal weights for the four classifiers were determined to be 4, 1, 1, and 4, respectively.

\begin{table}[]
\centering
\caption{List of parameter values for four machine learning classifiers.}
\label{table:parameter}
\scalebox{0.68}{
\begin{tabular}{ccc|ccc|ccc|ccc}
\hline
\textbf{Clsf.}            & \textbf{Para.}  & \textbf{Val.} & \textbf{Clsf.}            & \textbf{Para.}  & \textbf{Val.} & \textbf{Clsf.}            & \textbf{Para.}  & \textbf{Val.} & \textbf{Clsf.}            & \textbf{Para.}  & \textbf{Val.} \\ \hline
\multirow{4}{*}{\makecell{RF}} & max\_depth          & 10             & \multirow{3}{*}{KNN} & N\_neighbors       & 385            & \multirow{2}{*}{SVM} & kernel             & rbf            & \multirow{5}{*}{XGBoost} & eval\_metric       & mlogloss       \\
                               & n\_estimators       & 280            &                      & metric             & manhattan      &                      & gamma              & scale          &                          & max\_depth         & 9              \\
                               & min\_samples\_split & 5              &                      & weights            & distance       &                      &                    &                &                          & learning\_rate     & 0.01           \\
                               & min\_samples\_leaf  & 2              &                      &                    &                &                      &                    &                &                          & n\_estimators      & 95             \\
                               &                     &                &                      &                    &                &                      &                    &                &                          & colsample\_bytree  & 0.5            \\ \hline
\end{tabular}}
\end{table}

To address class imbalance in Phase II, we employed a conditional mixed sampling strategy to optimize the sample distribution. This included oversampling 65\% of bugs initially labeled as ``Minor,'' 15\% oversampling for ``Blocker'' and ``Critical'' bugs, 10\% undersampling for ``Major'' bugs, and 80\% oversampling for ``Trivial'' bugs. This strategy balanced the sample numbers across categories. Building on this, we used an Artificial Neural Network (ANN) with three hidden layers, each containing 32 neurons, trained over 86 epochs with a batch size of 64. To ensure the effectiveness and generalizability of the model, we adopted an unbiased sampling method, which can well preserve the distribution characteristics of the original data. Based on this sampling method, we divided the sampled dataset into a training set and a test set in a ratio of 8:2. It is worth mentioning that since sampling was performed in both phases, the number of samples for each project was only a few hundred on average after sampling. In this case, training the model for each project individually will result in insufficient data, which may affect the performance and generalization ability of the model. Therefore, when training the model, we chose to merge the data of all projects together for training, rather than training the model for each project separately. This global training strategy can make full use of limited data and improve the learning effect of the model.


\subsection{Evaluation Metrics}
Given that the bug priority change prediction in this study is divided into two phases—Phase I is a binary prediction, and Phase II is a multiclass prediction—different model evaluation metrics will be used for each phase. Specifically, the performance of the model in Phase I will be evaluated using \textit{Precision}, \textit{Recall}, and \textit{F1-Score}, which are standard metrics and do not require further elaboration. For Phase II, we will use \textit{F1-weighted} and \textit{F1-macro} as evaluation metrics, as detailed in Table \ref{table:metrics}. Given the class imbalance in this study, \textit{F1-weighted} and \textit{F1-macro} provide a comprehensive measure of the model’s overall performance across all classes.

\begin{table}[]
\centering
\caption{Evaluation Metrics of Phase II.}
\label{table:metrics}
\scalebox{0.80}{
\begin{tabular}{p{0.15\columnwidth} p{0.18\columnwidth} p{0.84\columnwidth} }
\hline
\textbf{Metric}      & \textbf{Formula} & \textbf{Description}                                                                               \\
\hline
\textit{F1-weighted} & $\sum_{i = 1}^{n}\left( \frac{s_{i}}{\sum_{j = 1}^{n}s_{j}}{F1}_{i} \right)$         & Weighted average of F1 scores for each class based on the sample size of each class. Here, \( n \) denotes the number of classes, \( s_i \) denotes the number of samples in the \( i \)-th class, and \( F1_i \) denotes the \textit{F1-score} of the \( i \)-th class.\\
\textit{F1-macro}    &  $\frac{1}{n}{\sum_{i = 1}^{n}{F1}_{i}}$        & Simple average of F1 scores for each class. \\  \hline                                          
\end{tabular}}
\end{table}

\subsection{Experimental Results}\label{sec_results}
This section addresses the three RQs formulated in Section \ref{sec:rq} through multiple ablation and comparative experiments. These experiments comprehensively evaluate the performance of the proposed bug priority change prediction method and verify the effectiveness and advantages of the bug fixing evolution features and class imbalance handling strategy. 
\subsubsection{Results for RQ1}
\label{sec_rq1}
To answer RQ1, we designed a series of ablation experiments for comparative analysis. Our data features are divided into two parts: basic features and bug fixing evolution features. Basic features encompass the basic features of bugs and text features. The class imbalance processing module is split into data processing and algorithm processing levels. Data processing measures include K-means undersampling (Phase I) and conditional mixed sampling (Phase II). Algorithm processing measures involve model ensembling (Phase I) and cost-sensitive learning (Phase II).
The ablation experiments aim to evaluate the effectiveness of each module by testing different combinations. The specific ablation experiment combinations are detailed in Table \ref{table:ablation}. In addition to the basic features of bugs, we combined the other three modules. It is worth noting that, due to the use of different data processing and algorithm processing measures at different phases, each row in Table \ref{table:ablation} represents a unique module combination method. The specific combinations include the models from both Phase I and Phase II. Before presenting the results, we clarify the meaning of each ablation combination shown in Table \ref{table:ablation}. Each acronym represents a unique combination of up to four modules: \textbf{Basic Features} (including basic bug attributes and text features), our proposed \textbf{Bug Fixing Evolution features (BFE)}, the \textbf{Data Processing} module for class imbalance \textbf{(D)}, and the \textbf{Algorithm Processing} module for class imbalance \textbf{(A)}. The specific combinations are defined as follows: 
\begin{itemize}
\item \textbf{Basic}: The baseline model, using only the Basic Features.
\item \textbf{BFE}: A model combining Basic Features with our proposed Bug Fixing Evolution features.
\item \textbf{BasicD} and \textbf{BasicA}: Models that augment the Basic model with either the Data Processing (D) or the Algorithm Processing (A) module, respectively. These processing strategies test the individual contribution of each imbalance handling module.
\item \textbf{BFED} and \textbf{BFEA}: Models that augment the BFE model with either the Data Processing (D) or the Algorithm Processing (A) module.
\item \textbf{BasicDA}: A model combining Basic Features with both imbalance handling modules (D and A), but excluding the BFE Features. This helps isolate the combined effect of the imbalance strategies.
\item \textbf{Our Method}: The complete proposed model, integrating all four modules (Basic Features, BFE, Data Processing, and Algorithm Processing).
\end{itemize}

\begin{table}[]
\centering
\caption{Combination of ablation experiments.}
\label{table:ablation}
\scalebox{0.80}{
\begin{tabular}{>{\centering\arraybackslash}p{0.18\columnwidth} >{\centering\arraybackslash}p{0.18\columnwidth} >{\centering\arraybackslash}p{0.22\columnwidth} >{\centering\arraybackslash}p{0.15\columnwidth} >{\centering\arraybackslash}p{0.16\columnwidth}}
\hline
\makecell{\textbf{Ablation} \\ \textbf{Combination}} & \makecell{\textbf{Basic Features}} & \makecell{\textbf{Bug Fixing} \\ \textbf{Evolution Features}} & \makecell{\textbf{Data} \\ \textbf{Processing}} & \makecell{\textbf{Algorithm} \\ \textbf{Processing}} \\ \hline
Basic          & \checkmark              &                               &                       &                \\
BFE            & \checkmark              & \checkmark                    &                       &                \\
BasicD         & \checkmark              &                               & \checkmark            &                \\
BasicA         & \checkmark              &                               &                       & \checkmark     \\
BFED           & \checkmark              & \checkmark                    & \checkmark            &                \\
BFEA           & \checkmark              & \checkmark                    &                       & \checkmark     \\
BasicDA        & \checkmark              &                               & \checkmark            & \checkmark     \\
Our Method     & \checkmark              & \checkmark                    & \checkmark            & \checkmark     \\ \hline
\end{tabular}}
\end{table}

To answer RQ1, we conducted a series of ablation experiments to evaluate the contribution of each component of our method. The performance of various model configurations, including our proposed full method and a heuristic \textbf{Baseline}, is presented in Table~\ref{table:ablationRes}. The evaluation is reported separately for Phase~I (\textit{Precision}, \textit{Recall}, \textit{F1-score}) and Phase~II (\textit{F1-weighted}, \textit{F1-macro}).

\textbf{For Phase I.} The results demonstrate a clear trade-off between \textit{Precision} and \textit{Recall} depending on the imbalance handling strategy. Models without data-level undersampling (e.g., \textbf{BFEA}, \textbf{BasicA}) achieve high \textit{Precision} (up to \textbf{0.878}) but exhibit extremely low \textit{Recall} (below 0.02). Conversely, models incorporating data processing (\textbf{BasicD}, \textbf{BFED}, \textbf{BasicDA}, \textbf{Our Method}) show a much more balanced performance, with substantially higher \textit{Recall} (above 0.79). Considering the comprehensiveness of \textit{F1-score}, which balances \textit{Precision} and \textit{Recall}, \textbf{Our Method} achieves the highest performance (\textbf{0.798}). This score is notably higher than the \textbf{Baseline} (0.553), the \textbf{Basic} model (0.031), and all other ablation variants.

\textbf{For Phase II.} The results consistently show performance gains from the inclusion of our proposed features and imbalance handling strategies. Comparing models with and without our bug fixing evolution features (\textbf{BFE} vs. \textbf{Basic}), the inclusion of these features yields a modest improvement in both \textit{F1-weighted} (0.634 vs. 0.624) and \textit{F1-macro} (0.527 vs. 0.516). The impact of the class imbalance handling strategies is more pronounced. Models incorporating data processing (\textbf{D}) or algorithm-level processing (\textbf{A}) consistently outperform the \textbf{Basic} and \textbf{BFE} models across both metrics. Ultimately, \textbf{Our Method}, which integrates all components, achieves the highest scores in both \textit{F1-weighted (0.712)} and \textit{F1-macro (0.613)}, surpassing all other configurations and the heuristic \textbf{Baseline} (0.367 and 0.312, respectively).

\begin{tcolorbox}[colback=gray!10!white, colframe=gray!75!black]
\textbf{Answer to RQ1:} Overall, our proposed bug fixing evolution features and class imbalance handling strategy substantially improved the model performance. Notably, the class imbalance handling strategy has a larger impact on performance than the bug fixing evolution features. Our proposed method achieves an \textit{F1-score} of 0.798 in Phase I, attains an \textit{F1-weighted} of 0.712 and an \textit{F1-macro} of 0.613 in Phase II.
\end{tcolorbox}

\begin{table}[]
\centering
\caption{Comparison of ablation experiment results.}
\label{table:ablationRes}
\scalebox{0.80}{
\begin{tabular}{ccccccccccc}
\hline
\textbf{Phase}             & \textbf{Metric}      & \textbf{Baseline}       & \textbf{Basic}  & \textbf{BFE}    & \textbf{BasicD} & \textbf{BasicA} & \textbf{BFED}   & \textbf{BFEA}            & \textbf{BasicDA} & \textbf{Our Method}          \\ \hline
\multirow{3}{*}{I} & \textit{Precision} & 0.553 & 0.840 & 0.847 & 0.729 & 0.852 & 0.757 & \textbf{0.878} & 0.748  & 0.776          \\
                  & \textit{Recall}    & 0.553 & 0.016 & 0.018 & 0.796 & 0.018 & 0.821 & 0.019          & 0.813  & \textbf{0.832} \\
                  & \textit{F1-score}  & 0.553 & 0.031 & 0.036 & 0.761 & 0.035 & 0.788 & 0.038          & 0.779  & \textbf{0.798} \\ \hline
\multirow{2}{*}{II} & \textit{F1-weighted}    & 0.367     & 0.624 & 0.634 & 0.681 & 0.665 & 0.701 & 0.682          & 0.692  & \textbf{0.712} \\
                  & \textit{F1-macro}      & 0.312      & 0.516 & 0.527 & 0.558 & 0.576 & 0.599 & 0.595          & 0.569  & \textbf{0.613} \\ \hline
\end{tabular}
}
\end{table}

\subsubsection{Results for RQ2}

\begin{figure*}
    \centering{\includegraphics[width=5.5in]{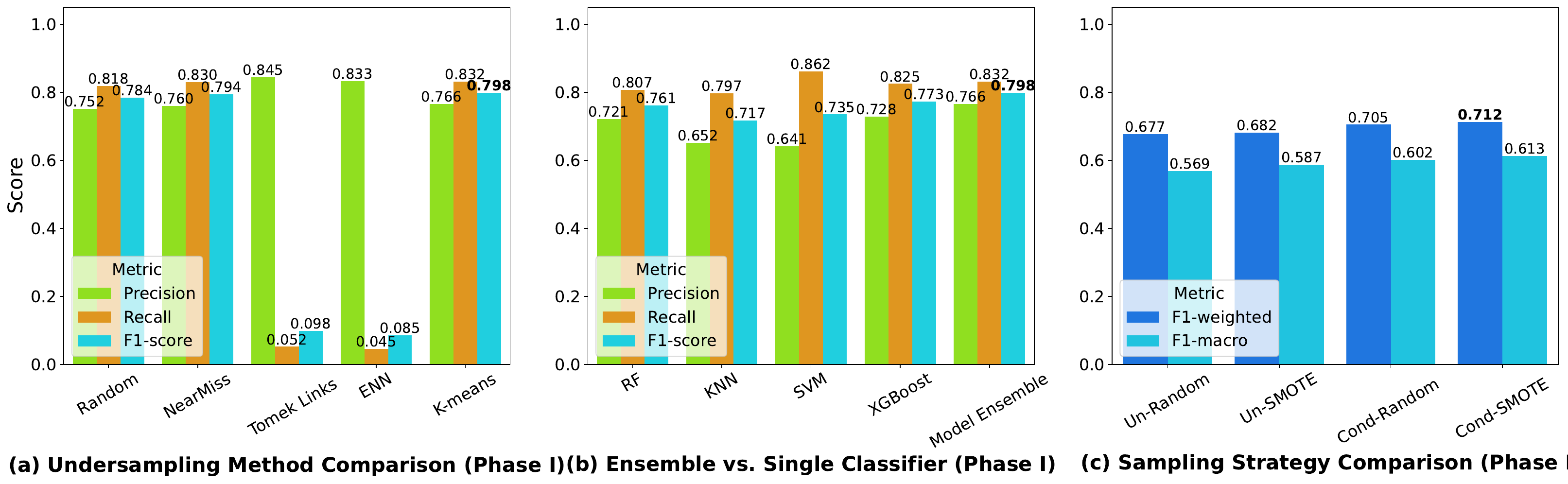}}
    \caption{Performance comparison of different class imbalance handling strategies for RQ2. (a) F1-scores of various undersampling methods in Phase I. (b) Comparison of the model ensemble against single classifiers in Phase I. (c) F1-weighted and F1-macro scores of different sampling strategies in Phase II. }
    \label{fig:rq2_performance_comparison}
\end{figure*}

To answer RQ2, we evaluated the effectiveness of our proposed class imbalance handling strategy by comparing it against several widely-used alternatives. The comparative results for both Phase~I and Phase~II are visualized in Fig.~\ref{fig:rq2_performance_comparison}. 
\textbf{For the data-level strategy in Phase I}, we compared our chosen K-means undersampling method against four other techniques. As shown in Fig.~\ref{fig:rq2_performance_comparison}(a), while the ``Tomek Links'' and ``ENN'' methods yield the highest \textit{Precision}, they suffer from extremely low \textit{Recall}, resulting in poor overall \textit{F1-scores}. In contrast, \textbf{K-means} and ``NearMiss'' achieve a much better balance. Notably, \textbf{K-means} demonstrates superior performance on the comprehensive \textit{F1-score} (\textbf{0.798}).
\textbf{For the algorithm-level strategy in Phase I}, we compared our model ensembling approach against the four individual base classifiers. The results in Fig.~\ref{fig:rq2_performance_comparison}(b) show that while the ``SVM'' model achieves the highest \textit{Recall}, our \textbf{Model Ensemble} method attains the highest \textit{Precision (0.766)} and, most importantly, the best overall \textit{F1-score (0.798)}, indicating its superior ability to balance precision and recall.
\textbf{For the sampling strategy in Phase II}, we compared our proposed conditional SMOTE mixed sampling (``Cond-SMOTE'') against three other mixed sampling variants. As illustrated in Fig.~\ref{fig:rq2_performance_comparison}(c), our \textbf{Cond-SMOTE} method consistently achieves the highest performance on both \textit{F1-weighted (0.712)} and \textit{F1-macro (0.613)}, outperforming all other strategies.

\begin{tcolorbox}[colback=gray!10!white, colframe=gray!75!black]
\textbf{Answer to RQ2:} In Phase I, at the data level, the K-means undersampling method outperformed both the random and ENN undersampling techniques; at the algorithm level, the performance of the model ensembling method we used exceeded that of individual models. In Phase II, the implemented conditional SMOTE mixed sampling method that combines conditional sampling with SMOTE oversampling outperformed other methods.
\end{tcolorbox}

\subsubsection{Results for RQ3}
\textbf{Performance of cross-project prediction models for bug priority changes.} Regarding RQ3, the 32 Apache projects selected for this study share many similarities, including their affiliation with the Apache Software Foundation (ASF) and similar project management styles and development processes. Additionally, some practitioners contribute to multiple projects simultaneously, and new projects often lack sufficient training data \cite{XiLoPaNa2016}. Therefore, it is essential to test the predictive performance of the model on each project. All data come from these 32 projects. To test the predictive performance, the overall method framework follows the method described in Fig. \ref{fig:PredMethod}, with the primary difference being the division of training and testing sets. For instance, to test the predictive performance on the Kafka project, the bug data from Kafka serve as the testing set, while the data from the remaining 31 projects form the training set. Table \ref{table:CrossProject} shows the prediction performance metrics for the Phase I and Phase II models across projects.

To answer RQ3, Table \ref{table:CrossProject} lists the corresponding metrics for Phase I (\textit{Precision}, \textit{Recall}, and \textit{F1-score}) and Phase II (\textit{F1-weighted} and \textit{F1-macro}) for each project. Light grey and dark grey cells represent the minimum and maximum values for that metric, respectively. Descriptive statistics are also provided at the bottom of the table, including the minimum, maximum, first quartile, third quartile, mean, and median for each metric. As shown in Table \ref{table:CrossProject}, there are large differences in the predictive performance of models for different projects in terms of the Phase I metrics (\textit{Precision}, \textit{Recall}, and \textit{F1-score}). For example, the \textit{F1-score} of project Ambari is only 0.378, while the \textit{F1-score} of project Guacamole is 0.928. According to Table \ref{table:proportion}, the priority change rate of project Ambari is 1.7\%, while the priority change rate of project Guacamole is 24.4\%, which respectively represent the lowest and highest priority change rates among the 32 projects. 
For the Phase II metrics (\textit{F1-weighted} and \textit{F1-macro}), Table \ref{table:CrossProject} demonstrates variations in model performance across projects. The \textit{F1-weighted} is generally higher than the \textit{F1-macro}, indicating varied predictive performance across different priorities. Specifically, the model performed best on project Flink  with \textit{F1-weighted} and \textit{F1-macro} of 0.789 and 0.713, respectively, while it performed worst on project Geode with \textit{F1-weighted} and \textit{F1-macro} of 0.623 and 0.512, respectively. For most projects, the \textit{F1-score} generally falls within the range of 0.7 to 0.9, while the \textit{F1-score} achieved by our method is 0.79. In Phase II, the \textit{F1-weighted} for the majority of projects lies between 0.6 and 0.8, with our method achieving an \textit{F1-weighted} of 0.7. Similarly, the \textit{F1-macro} for most projects ranges from 0.5 to 0.7, and our method attains an \textit{F1-macro} of 0.61. Overall, the performance of the bug priority change prediction model varies by project, but for most projects, the model matches the predictive ability of models trained on all project data.

\begin{table}[]
\centering
\caption{Model performance across projects.}
\label{table:CrossProject}
\scalebox{0.63}{
\begin{tabular}{cccccccccccc}
\hline
\multirow{2}{*}{\textbf{Project}} & \multicolumn{3}{c|}{\textbf{Phase I}}   & \multicolumn{2}{c|}{\textbf{Phase II}}               & \multirow{2}{*}{\textbf{Project}} & \multicolumn{3}{c|}{\textbf{Phase I}}    & \multicolumn{2}{c}{\textbf{Phase II}} \\ \cline{2-6} \cline{8-12}
& \textit{\textbf{Precision}} & \textit{\textbf{Recall}} & \multicolumn{1}{c|}{\textit{\textbf{F1-score}}} & \textit{\textbf{F1-weighted}} & \multicolumn{1}{c|}{\textit{\textbf{F1-macro}}}                   &                          & \textit{\textbf{Precision}} & \textit{\textbf{Recall}} & \multicolumn{1}{c|}{\textit{\textbf{F1-score}}} & \textit{\textbf{F1-weighted}}   & \textit{\textbf{F1-macro}}   \\ \hline
ActiveMQ                 & 0.721 & 0.658 & 0.688   & 0.742      & \multicolumn{1}{c|}{0.618} & Ignite                   & 0.722 & 0.700 & 0.711   & 0.742        & 0.618     \\
Ambari                   & \cellcolor[HTML]{D9D9D9}\textbf{0.390} & \cellcolor[HTML]{D9D9D9}\textbf{0.367} & \cellcolor[HTML]{D9D9D9}\textbf{0.378}   & 0.652      & \multicolumn{1}{c|}{0.547} & Impala                   & 0.784 & 0.788 & 0.786   & 0.652        & 0.547     \\
Arrow                    & 0.637 & 0.583 & 0.609   & 0.651      & \multicolumn{1}{c|}{0.554} & Jackrabbit-Oak           & 0.773 & 0.713 & 0.742   & 0.651        & 0.554     \\
Axis2                    & 0.751 & 0.868 & 0.805   & 0.662      & \multicolumn{1}{c|}{0.562} & Kafka                    & 0.663 & 0.781 & 0.717   & 0.662        & 0.562     \\
Camel                    & 0.832 & 0.924 & 0.876   & 0.757      & \multicolumn{1}{c|}{0.638} & Lucene                   & 0.713 & 0.697 & 0.705   & 0.757        & 0.638     \\
CloudStack               & 0.669 & 0.773 & 0.717   & 0.732      & \multicolumn{1}{c|}{0.628} & Mesos                    & 0.701 & 0.789 & 0.742   & 0.732        & 0.628     \\
Cordova                  & 0.750 & 0.928 & 0.829   & 0.741      & \multicolumn{1}{c|}{0.613} & NetBeans                 & 0.655 & 0.706 & 0.680   & 0.741        & 0.613     \\
Drill                    & 0.748 & 0.844 & 0.794   & 0.746      & \multicolumn{1}{c|}{0.609} & NiFi                     & 0.676 & 0.607 & 0.639   & 0.746        & 0.609     \\
Flink                    & 0.779 & 0.909 & 0.839   & 0.790      & \multicolumn{1}{c|}{0.714} & OFBiz                    & 0.704 & 0.725 & 0.714   & \cellcolor[HTML]{ADADAD}\textbf{0.790} & \cellcolor[HTML]{ADADAD}\textbf{0.714}     \\ 
Geode                    & 0.846 & 0.427 & 0.567   & \cellcolor[HTML]{D9D9D9}\textbf{0.624}      & \multicolumn{1}{c|}{\cellcolor[HTML]{D9D9D9}\textbf{0.513}} & Ozone                    & 0.686 & 0.777 & 0.729   & 0.624        & 0.513     \\ 
Groovy                   & 0.767 & 0.775 & 0.771   & 0.721      & \multicolumn{1}{c|}{0.612} & Qpid                     & 0.718 & 0.607 & 0.657   & 0.721        & 0.612     \\ 
Guacamole                & \cellcolor[HTML]{ADADAD}\textbf{0.881} & \cellcolor[HTML]{ADADAD}\textbf{0.981} & \cellcolor[HTML]{ADADAD}\textbf{0.928}   & 0.769      & \multicolumn{1}{c|}{0.650} & Solr                     & 0.684 & 0.619 & 0.650   & 0.769        & 0.650     \\ 
Hadoop                   & 0.810 & 0.916 & 0.860   & 0.736      & \multicolumn{1}{c|}{0.611} & Spark                    & 0.722 & 0.778 & 0.749   & 0.736        & 0.611     \\ 
HBase                    & 0.739 & 0.794 & 0.766   & 0.699      & \multicolumn{1}{c|}{0.589} & Thrift                   & 0.717 & 0.882 & 0.791   & 0.699        & 0.589     \\ 
Hive                     & 0.641 & 0.465 & 0.539   & 0.647      & \multicolumn{1}{c|}{0.539} & Traffic Server           & 0.701 & 0.716 & 0.709   & 0.647        & 0.539     \\ 
Hudi                     & 0.778 & 0.863 & 0.818   & 0.650      & \multicolumn{1}{c|}{0.571} & Wicket                   & 0.785 & 0.843 & 0.813   & 0.650        & 0.571     \\ \hline 
\multicolumn{12}{c}{\textbf{Descriptive Statistics}} \\ \hline
\textbf{Min}             & 0.390    & 0.367 & 0.378   & 0.624      & \multicolumn{1}{c|}{0.513} & \textbf{Max}             & 0.881    & 0.981 & 0.928   & 0.790        & 0.714     \\
\textbf{1st Qu.}         & 0.686    & 0.699 & 0.686   & 0.652      & \multicolumn{1}{c|}{0.554} & \textbf{3rd Qu.}         & 0.778    & 0.865 & 0.810   & 0.744        & 0.628     \\
\textbf{Mean}            & 0.729    & 0.762 & 0.741   & 0.705      & \multicolumn{1}{c|}{0.597} & \textbf{Median}          & 0.731    & 0.776 & 0.742   & 0.727        & 0.611           \\ \hline
\end{tabular}
}
\end{table}

\begin{figure*}
    \centering{\includegraphics[width=5.5in]{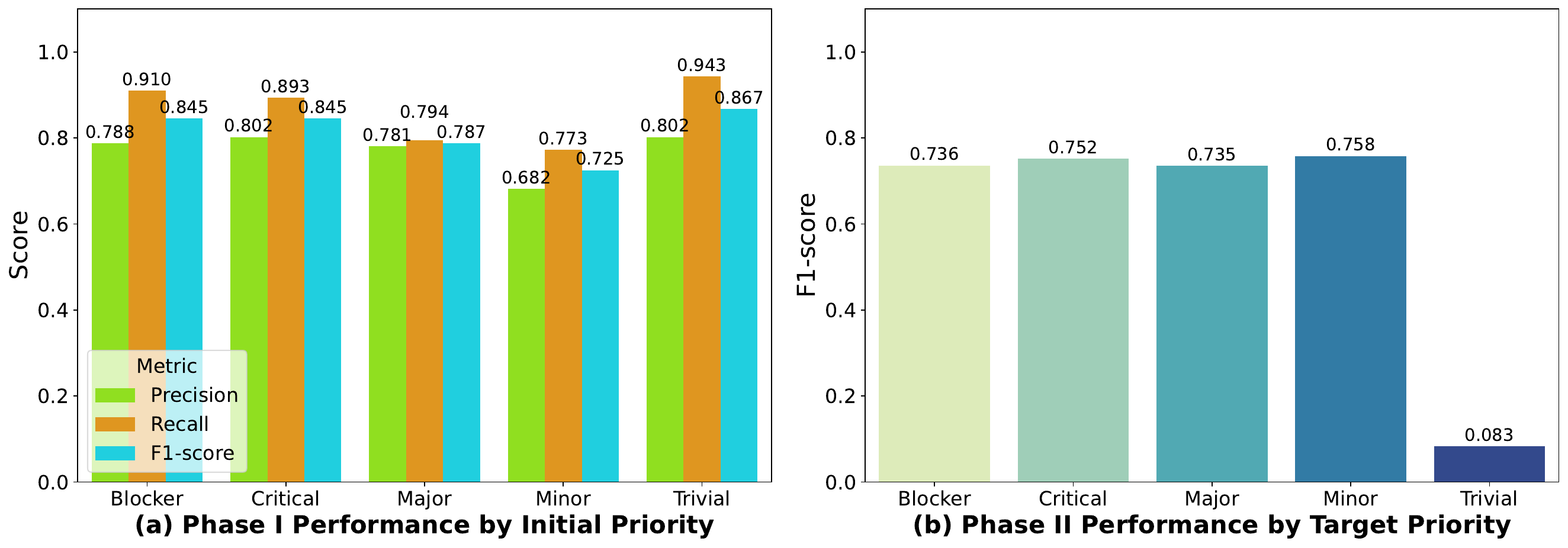}}
    \caption{Model performance across different priority levels for Phase I and Phase II. (a) shows the Precision, Recall, and F1-score for the Phase I model based on the bug's initial priority. (b) shows the F1-score for the Phase II model for each target priority.}
    \label{fig:rq3_priority_performance}
\end{figure*}

\textbf{Performance Across Different Priority Levels.} To further explore the performance of our prediction models at different priorities, we visualized the results for both Phase~I and Phase~II in Fig.~\ref{fig:rq3_priority_performance}. For Phase~I, as shown in Fig.~\ref{fig:rq3_priority_performance}(a), the model demonstrates robust and well-balanced performance across most initial priority levels. Specifically, it achieves excellent \textit{F1-scores} for ``Blocker'' (0.845), ``Critical'' (0.845), and ``Trivial'' (0.867) bugs. The performance on ``Major'' (0.787) and ``Minor'' (0.725) priorities is slightly lower than that on the other priorities but remains strong. The grouped bars also illustrate that this performance is generally driven by high \textit{Recall} values, particularly for the ``Trivial'' priority (0.943). For Phase~II, Fig.~\ref{fig:rq3_priority_performance}(b) reveals a contrasting pattern. The model exhibits remarkably consistent and strong performance for the four highest target priorities: ``Blocker'' (0.736), ``Critical'' (0.752), ``Major'' (0.735), and ``Minor'' (0.758). However, in stark contrast, the model performs extremely poorly when predicting the ``Trivial'' priority, with the \textit{F1-score} dropping dramatically to just 0.083.

\begin{tcolorbox}[colback=gray!10!white, colframe=gray!75!black]
\textbf{Answer to RQ3:} For cross-project prediction model performance, in Phase I although the \textit{F1-score} varies substantially depending on the project, ranging from 0.372 (Ambari) to 0.928 (Guacamole), 75\% (24/32) of the projects achieve an \textit{F1-score} greater than 0.70 (Table \ref{table:CrossProject}); in Phase II, the \textit{F1-weighted} and \textit{F1-macro} range from 0.623 and 0.512 (Geode) to 0.789 and 0.713 (Flink) (Table \ref{table:CrossProject}), respectively. Regarding different priority levels, the prediction model in Phase I performs well for ``Blocker'', ``Critical'', and ``Trivial'' bugs with \textit{F1-scores} greater than 0.84, and decently for ``Major'' and ``Minor'' bugs with \textit{F1-scores} greater than 0.72 (Fig. \ref{fig:rq3_priority_performance}); in Phase II, the model performs poorly for ``Trivial'' bugs, but achieves consistently decent performance for bugs of other priority levels with \textit{F1-scores} falling in the range (0.73, 0.76) (Fig. \ref{fig:rq3_priority_performance}).
\end{tcolorbox}

\section{Discussion}\label{sec_discussion}

\subsection{Interpretation of Study Results}
Our experimental results, presented in Section \ref{sec_results}, not only quantify the performance of our method but also reveal important insights into the nature of bug priority change prediction. This section provides a deeper interpretation of the findings for each research question.

\textbf{Granger Causality Analysis.} To further investigate the mechanisms underlying bug priority changes, we conducted Granger causality analysis on several bug evolution features, including \textit{bug comment frequency}, \textit{commenter frequency}, \textit{comment length frequency}, \textit{history change frequency}, \textit{history participant frequency}, and \textit{history field change frequency}. We randomly sampled 1,000 priority change events and performed the following steps:

\begin{enumerate}
    \item \textbf{Pre-event window selection}: For each event, we defined the observation window $W_{\text{pre}}$, starting from the bug reporting time (or the previous priority change, if any) and ending at the current priority change.
    \item \textbf{Data collection}: Within $W_{\text{pre}}$, we collected the values of the above evolution features.
    \item \textbf{Lead-label construction}: We generated a lead-label time series for each event by labeling the $H$ time slices before the change as 1 and the others as 0, ensuring sufficient variance for fitting the VAR model ($H=2$ in our analysis).
    \item \textbf{Time series normalization}: Because bug lifecycles vary greatly (ranging from minutes to weeks), we rescaled each sequence into a fixed number of relative time slices ($N=10$).
    \item \textbf{Lag order and Granger testing}: We set the maximum lag order to $L_{\max}=3$ and performed Granger causality tests for each feature.
    \item \textbf{Multiple testing correction}: We applied FDR correction (Benjamini--Hochberg) to all results, controlling the false discovery rate at $q < 0.05$.
    \item \textbf{Result aggregation}: For each feature, we reported the percentage of significant cases, median lag, median F-statistic, and median adjusted q-value.
\end{enumerate}

The results are shown in Table~\ref{tab:granger}. For example, \textit{comment frequency} was significant in 34.4\% of the events after FDR correction ($q < 0.05$), with a median lag of 3, suggesting that systematic increases in comment activity during the 1--3 preceding time slices strongly predict upcoming priority changes. Its median F-statistic was 4.11, reflecting a moderate predictive relationship with temporal precedence (with $F < 2$ considered weak, $2$--$5$ moderate, and $>5$ moderate-to-strong). In contrast, features such as \textit{commenter frequency} and \textit{history participant frequency} showed much lower significance rates, suggesting weaker predictive power. Overall, Granger causality analysis provides a temporal and quasi-causal perspective that complements our predictive modeling, helping researchers and practitioners better understand which aspects of bug activity are most informative for predicting priority changes. We emphasize that Granger causality captures temporal precedence in prediction rather than strict causation; nevertheless, it offers actionable insights into the dynamics of bug management workflows.

\begin{table}[ht]
\centering
\caption{Results of Granger causality analysis on bug evolution features.}
\label{tab:granger}
\scalebox{0.80}{
\begin{tabular}{lcccc}
\toprule
\textbf{Feature} & \textbf{\% Significant ($q<0.05$)} & \textbf{Median Lag} & \textbf{Median F} & \textbf{Median $q$} \\
\midrule
comment frequency           & 34.4 & 3 & 4.11 & 0.021 \\
commenter frequency         & 14.2 & 1 & 1.22 & 0.057 \\
comment length frequency    & 21.1 & 1 & 2.57 & 0.045 \\
history change frequency    & 25.3 & 3 & 3.35 & 0.027 \\
history participant frequency & 9.6 & 1 & 1.18 & 0.069 \\
history field change frequency & 11.2 & 2 & 1.89 & 0.048 \\
\bottomrule
\end{tabular}
}
\end{table}

\textbf{The Power of Dynamic Features and Imbalance Handling.} The ablation experiments (Table \ref{table:ablationRes}) clearly demonstrate that both our novel bug fixing evolution features (BFE) and the class imbalance handling strategy substantially improve model performance over a basic model. For Phase I, the results highlight a critical trade-off. Models without data-level undersampling (e.g., BFEA) achieve high Precision but near-zero Recall, rendering them practically useless for identifying the rare ``change'' events. This confirms that for the highly imbalanced binary prediction task, a robust imbalance handling strategy is not just beneficial, but essential. The dramatic increase in F1-score from BFE (0.036) to Our Method (0.798) is overwhelmingly driven by the undersampling and ensembling techniques. This suggests that in the initial prediction phase, correctly balancing the data distribution is the single most important factor for building a useful model. For Phase II, the results are nuanced. The inclusion of evolution features (BFE vs. Basic) provides a modest but consistent performance gain. This indicates that the dynamic signals captured by these features do contain valuable predictive information about how a priority will change. Nevertheless, the improvement from imbalance handling strategies (BasicD, BasicA) is significantly more pronounced. This implies that even in the more complex multi-class scenario, accurately representing the minority classes is a primary challenge.

\textbf{The Superiority of Chosen Imbalance Handling Techniques.} Our comparative experiments for RQ2 (Fig. \ref{fig:rq2_performance_comparison}) further justify our specific methodological choices and reveal insights into why they are effective for this task.
In Phase I, the superiority of K-means undersampling (Fig. \ref{fig:rq2_performance_comparison}(a)) over methods like Tomek Links or ENN is particularly telling. Tomek Links and ENN, which focus on cleaning class boundaries, remove only a few majority class samples, failing to resolve the severe class imbalance. This results in high Precision but catastrophically low Recall. K-means, by clustering the majority class and selecting representative centroids, effectively reduces the class imbalance while preserving the underlying distributional structure of the data, leading to a much better balance between Precision and Recall. Similarly, the success of Model Ensembling (Fig. \ref{fig:rq2_performance_comparison}(b)) over any single classifier highlights the complexity of the feature space; no single model can perfectly capture all the patterns, and by combining their probabilistic outputs, we create a more robust and generalized decision boundary.
In Phase II, the outperformance of Conditional SMOTE mixed sampling (Cond-SMOTE) (Fig. \ref{fig:rq2_performance_comparison}(c)) is also significant. The clear advantage of ``conditional'' sampling over ``unconditional'' variants demonstrates that the initial priority level is a powerful contextual factor. By preserving the original transition probabilities for each initial priority during resampling, our Cond-SMOTE method generates more realistic synthetic samples, avoiding the introduction of ``noise'' that an unconditional approach might create. This allows the model to learn from a dataset that is both quantitatively balanced and qualitatively authentic.

\textbf{Analysis of Cross-Project Performance Variation.} To investigate the underlying causes of the cross-project performance variations observed in our evaluation of RQ3, we conducted a Spearman correlation analysis. This analysis quantitatively examined the relationship between project characteristics and performance metrics of our models. The results are presented in Table~\ref{tab:corr_simple}, where shaded cells indicate a statistically significant correlation ($p<0.05$) between a project characteristic and a performance metric. The analysis yielded several key insights. First, a strong and statistically significant positive correlation exists between a project's priority change probability (i.e., Priority Change Prob.) and the performance of our Phase I model (F1-score: $\rho = 0.77, p<0.001$). This finding provides robust empirical evidence for our central hypothesis: projects with a higher historical incidence of priority changes offer a richer and more balanced training set for the binary classification task. Balanced training data mitigate the dominance of the majority class (no change), largely explaining the source of the performance variance in Phase I. Furthermore, the analysis identified crucial factors influencing the more complex multi-class prediction task in Phase II. Model performance, particularly the F1-macro metric, is significantly and positively correlated with both project age (Age (years)) and the number of commits (\#Revisions). As the characteristics for project maturity and the richness of historical development activities, these results suggest that our Phase II model benefits from the extensive and diverse data patterns present in long-lived, more active projects. Such projects likely possess more established and consistent bug resolution processes, providing clearer signals for the model to learn from. Conversely, it is equally insightful that simple scale metrics, such as the number of committers (\#Committers) or the total number of bugs in JIRA (\#Bugs in JIRA), show no significant correlation with model performance. This strongly suggests that our model's effectiveness is less dependent on the size of a project, and is instead more deeply rooted in its dynamic and evolutionary characteristics. 

\textbf{Analysis of Model Robustness across Languages and Domains.} In addition to the Spearman correlation analysis, we further assessed the robustness of our model's performance by conducting a series of Mann-Whitney U tests, grouping projects by both programming language and application domain. 
First, projects were categorized by their implementation language into three groups: ``Java-only'', ``Java-centric (multi-language)'', and ``Non-Java'' projects (detailed grouping is available in our replication package \cite{dataset}). The pairwise Mann-Whitney U tests performed across these groups revealed no statistically significant differences in any of the performance metrics after applying a Bonferroni correction (all corrected $p > 0.05$). Furthermore, the calculated effect sizes (Rank-Biserial Correlation, $r$) for these comparisons were consistently in the trivial-to-small range (all $\lvert r \rvert < 0.30$), indicating a lack of practically meaningful differences.
Second, projects were classified into four major application domains (as visualized in Fig.~\ref{fig:projects_bubble} in Section \ref{process_data_processing}). Again, the Mann-Whitney U tests, after p-value correction, demonstrated that the model's performance was largely consistent across these domains. The vast majority of pairwise comparisons yielded no statistically significant differences (corrected $p > 0.05$). However, our effect size analysis revealed a more nuanced picture: while most effect sizes were negligible, a few specific pairwise comparisons exhibited large effect sizes (up to $\lvert r \rvert = 1.00$). This suggests that while not statistically significant with our current sample size, some domain-specific characteristics may have a practical impact on our models' performance and warrant future investigation.
Collectively, these tests provide strong evidence that our model is largely language-agnostic. While the performance is also generally consistent across application domains (most corrected $p > 0.05$), the presence of large effect sizes in some comparisons suggests that certain domain-specific factors may still influence the performance, warranting further investigation in larger-scale studies.

\begin{table}[ht]
\centering
\caption{Spearman correlation tests between project characteristics and model performance metrics.}
\label{tab:corr_simple}
\scalebox{0.85}{
\begin{tabular}{lccccc}
\toprule
\textbf{Project Characteristic} & {\textbf{Precision}} & {\textbf{Recall}} & {\textbf{F1-score}} & {\textbf{F1-weighted}} & {\textbf{F1-macro}} \\
\midrule
Priority Change Prob.        & \cellcolor[HTML]{ADADAD}0.42, 0.01  & \cellcolor[HTML]{808080}0.79, \textless 0.01 & \cellcolor[HTML]{808080}0.77, \textless 0.01 & 0.26, 0.15         & 0.26, 0.15        \\
Age (years)                  & 0.20, 0.26         & 0.09, 0.59         & 0.12, 0.53         & 0.24, 0.18        & \cellcolor[HTML]{ADADAD}0.35, 0.04   \\
\#Revisions                  & 0.07, 0.71         & -0.03, 0.87       & 0.02, 0.92         & \cellcolor[HTML]{ADADAD}0.37, 0.03   & \cellcolor[HTML]{ADADAD}0.49, \textless 0.01  \\
\#Committers                 & -0.13, 0.48        & 0.22, 0.21         & 0.16, 0.38         & 0.15, 0.42        & 0.13, 0.48        \\
\#Bugs in JIRA               & -0.11, 0.55        & -0.09, 0.61        & -0.09, 0.61        & 0.15, 0.40         & 0.10, 0.58         \\
\#Bugs with Priority Changes & 0.05, 0.79         & 0.34, 0.06        & 0.31, 0.08        & \cellcolor[HTML]{ADADAD}0.35, 0.04   & 0.29, 0.10        \\
\bottomrule
\end{tabular}
}
\end{table}

\textbf{Analysis of Poor Performance on the Trivial Priority Level.} A notable finding in our evaluation of RQ3 is the model's exceptionally poor performance in predicting changes to the ``Trivial'' priority level, achieving an F1-score of only 0.0831. To investigate the underlying reasons for this result, we conducted a detailed analysis focusing on data distribution and feature signals. First, the \textbf{``Trivial'' class suffers from severe data sparsity and internal imbalance}. As shown in Table~\ref{table:PCDistribution} in Section \ref{sec:imblance_2}, out of 21,215 total change records, only 369 originated from a ``Trivial'' priority, accounting for less than 1.8\% of the dataset. This extreme rarity means that the model lacks sufficient examples from which to learn. Although our methodology included an aggressive 80\% oversampling for this class, this could not fundamentally overcome the scarcity of unique, original samples. Furthermore, within these 369 instances, the distribution is also highly skewed: 72.4\% (267 samples) change to ``Minor'', while the number of changes to other classes is negligible. This dual imbalance makes it exceedingly difficult for the model to learn reliable patterns for transitions from ``Trivial'' to non-``Minor'' priorities. Second, \textbf{bugs originating as ``Trivial'' generally exhibit weaker predictive signals}. Our analysis of key textual features, presented in Table~\ref{tab:feature_means}, reveals that ``Trivial'' bugs have, on average, a significantly shorter \textit{Summary Length}, \textit{Description Length}, and \textit{Total Comment Length} compared to all four other priority levels. Mann-Whitney U tests confirmed that these differences are statistically significant (except for the p-value of 0.082 for Trivial vs. Minor on \textit{Total Comment Length}, all other p-values are less than 0.05). This suggests that ``Trivial'' bugs typically receive less attention, generate less discussion, and are documented with less detail. This lack of strong signals results in low separability in the feature space, making it challenging for the model to differentiate them from other classes based on the available, weaker evidence. In summary, the poor performance on the ``Trivial'' priority is a consequence of the combined effects of \textbf{data sparsity} and \textbf{weak predictive signals}. This highlights an inherent challenge for the model when dealing with extremely rare classes that also lack distinct behavioral characteristics.

\begin{table}[ht]
\centering
\caption{Average length (in bytes) of bug summary, description, and total comments for different priorities, with Mann-Whitney U test p-values comparing each priority level against Trivial.}
\label{tab:feature_means}
\scalebox{0.75}{ 
\begin{tabular}{lcccccc}
\toprule
\textbf{Priority Level} & {\textbf{Summary Length}} & \textbf{p-value} & {\textbf{Description Length}} & \textbf{p-value} & {\textbf{Total Comment Length}} & \textbf{p-value} \\
\cmidrule(lr){2-3} \cmidrule(lr){4-5} \cmidrule(lr){6-7} 
 & {Mean} & {vs Trivial} & {Mean} & {vs Trivial} & {Mean} & {vs Trivial} \\
\midrule
Blocker & 64.32 & $<$0.001 & 2870.27 & $<$0.001 & 2892.60 & $<$0.001 \\
Critical & 65.77 & $<$0.001 & 3032.27 & $<$0.001 & 2609.97 & 0.039 \\
Major & 64.14 & $<$0.001 & 2332.80 & $<$0.001 & 2261.22 & 0.046 \\
Minor & 64.53 & $<$0.001 & 1569.24 & $<$0.001 & 2045.42 & 0.082 \\
Trivial & \textbf{57.94} & {-} &  \textbf{778.38} & {-} & \textbf{1369.28} & {-} \\
\bottomrule
\end{tabular}
}
\end{table}

\subsection{Implications for Practitioners}
Our findings offer several practical implications for software development teams and project managers seeking to improve their bug triage and resolution workflows. The proposed two-phase bug priority change prediction model is not merely a theoretical construct; it is designed to be integrated into existing issue tracking systems (ITS) like JIRA or Bugzilla to provide actionable, real-time decision support.

\textbf{Integration into Development Workflows.} For our model to deliver practical value, it must be seamlessly integrated into the existing development ecosystem. We propose two primary deployment architectures that cater to different operational needs and technical environments, moving beyond manual checks to a fully automated monitoring system. \textbf{(1) Periodic Batch Processing via a Backend Service.} This architecture involves deploying the prediction model as a standalone, asynchronous service that runs independently of the main ITS application. The service can be implemented as a scheduled task (e.g., a Cron job) that executes at regular intervals (e.g., hourly or nightly). During each run, it queries the ITS database or API for a list of active bugs that have been updated since the last check. For each of these bugs, it extracts the latest feature values, runs the Phase II prediction, and if the predicted priority differs from the current one, it can trigger a notification or update the bug report via the ITS API. This way is ideal for large-scale projects where real-time prediction is not critical but a regular, systematic review of bug priorities is desired. It is particularly useful for generating daily or weekly ``priority health reports'' for project managers. \textbf{(2) Real-Time Prediction via Event-Driven Integration.} This architecture provides immediate feedback by tightly coupling the prediction model with the ITS through plugins or webhooks. The model is wrapped in an API and deployed as a service. The ITS (e.g., JIRA, Bugzilla) is then configured to send an HTTP request (a webhook) to this service whenever a relevant event occurs. The service receives the event payload (containing bug data), performs the prediction in real-time, and can immediately post a result back to the ITS. The following two events can trigger corresponding actions. 
\textbf{Bug Creation:} When a new bug report is submitted, a webhook is sent. The service invokes the Phase I model to predict the stability of its priority. If predicted to be unstable, the service uses the ITS API to automatically add a Priority-Unstable label to the new bug. \textbf{Bug Update} (e.g., Comment Added, Field Changed): When a significant update occurs, another webhook is sent. The service then invokes the Phase II model to perform a re-assessment. If a priority change is recommended, it can post an automated, private comment suggesting the change to the assignee. By offering these distinct integration pathways, our method can be flexibly adapted to the specific technical constraints and workflow preferences of different software development teams.

\textbf{The timing of priority changes.} A crucial aspect of our model's design is that it deliberately does not predict a future timestamp for when a priority will change. Instead, our two-phase method offers distinct temporal decision support. Phase I acts as a strategic ``Early Warning System'' at bug creation. Its purpose is not to forecast a precise time but to identify which bug priorities are inherently volatile and require proactive monitoring throughout their lifecycle. Phase II, in contrast, serves as a tactical ``Just-in-Time Decision Point''. It is triggered by real-time events (e.g., a new comment of a bug) and provides a recommendation to adjust the priority immediately. In this sense, the timing of the prediction is the timing of the decision; the model does not forecast a future change but rather determines that a change is justified now.

\textbf{Translating Model Predictions into Actionable Insights.}
Once integrated, the model's outputs from each phase can be translated into a series of concrete, actionable insights that enhance the proactiveness and efficiency of project management. 
\textbf{Phase I as an ``Early Warning System'':} The binary prediction from Phase~I provides an invaluable early warning signal at the moment of bug creation. Practitioners can leverage this signal to \textbf{Automatically Flag Potential Risks:} When the model predicts that a new bug's priority is likely to change, the system can automatically apply a label such as Priority-Unstable or Needs-Re-evaluation. This flag serves as a cue for triage managers to give the bug additional scrutiny during sprint planning or to assign it to a more senior developer for a precise secondary assessment.
\textbf{Phase II as a ``Just-in-Time Decision Support System'':} The multi-class prediction from Phase~II offers just-in-time, data-driven recommendations for priority adjustments during the bug's lifecycle. Teams can utilize these recommendations in several ways. \textbf{Trigger Proactive Resource Re-allocation:} If a bug initially marked as \texttt{Minor} is later predicted to escalate to \texttt{Critical} following a surge in activity, an automated notification can be triggered. This allows the team to proactively intervene by assigning senior resources before the issue becomes critical, thus shifting from a reactive to a proactive manner. \textbf{Provide Data-Driven Justification for Changes:} During sprint planning meetings, a developer can cite the model's prediction to support a proposal for reprioritization. For instance, they might state, ``\textit{Based on the recent comment frequency and historical patterns, the model suggests elevating this bug's priority to Critical}'', providing an objective basis for the decision. \textbf{Enable Dynamic Risk Monitoring:} The model's predictions can be integrated into project management dashboards like JIRA. A dedicated widget could display the ``Top 5 Predicted-to-Escalate Bugs'' in real-time, giving managers an at-a-glance view of potentially volatile bugs within the project. Crucially, we emphasize that the model is designed as a decision-support tool, not a decision replacement. Its predictions should be interpreted as strong, evidence-based recommendations derived from historical data. The final authority on any priority change must always rest with the developers, who can synthesize the model's output with their indispensable contextual knowledge of current project goals, client commitments, and team capacity.

\textbf{Feasibility and Cost of Feature Extraction.} A critical factor for the feasibility of any real-time prediction system is the computational cost of feature extraction. We argue that for our proposed model, this cost is minimal and well-suited for both periodic and event-driven deployment scenarios.
The majority of the features used by our model, including our novel evolutionary features, are derived from structured or semi-structured metadata that is already indexed and efficiently accessible in modern Issue Tracking Systems (ITS). These features can be categorized into two types: \textbf{Directly Available Attributes.} Many features, such as the number of comments, number of attachments, and the bug reporter's ID, are typically available directly as fields in the bug report object returned by a single API call (e.g., a \texttt{GET /rest/api/2/issue/{issueIdOrKey}} request in JIRA). The cost of extracting these is negligible. \textbf{Computed Historical Features.} More complex evolutionary features, such as the rate of change in comment frequency or assignment history, require retrieving a list of historical events. While this may involve a slightly more expensive API call (e.g., to the changelog endpoint), this operation is still highly optimized by the ITS backend. Once the history is retrieved, the subsequent computation - such as counting events within a time window or calculating deltas - is a lightweight, in-memory operation that takes only milliseconds.
Crucially, our approach avoids computationally expensive operations like natural language processing (NLP) on the entire bug history during every real-time prediction. The deep text features from RoBERTa are typically extracted only once at bug creation or during periodic updates, not in a synchronous, blocking manner. To this end, we randomly sampled text features from 1,000 records and extracted their feature representations using RoBERTa, with an average time of 180.58 ms (CPU: Intel i7-11800H, GPU: RTX 3060).
Similarly, we also randomly sampled 1,000 records each for Phase I and Phase II, with average model prediction time of 8.22 ms and 10.61 ms per record, respectively (hardware configuration as above). This ensures that the integration of our model does not introduce any noticeable delay into the interactive workflows of developers, making it a highly feasible solution for continuous integration and development environments. 

\textbf{The Strategic Trade-off between Timeliness and Accuracy.} Our two-phase method embraces the inherent trade-off between the timeliness of a prediction (Phase I) and the accuracy achievable with more complete information (Phase II). The primary value of the Phase I model is its immediacy. It provides a prediction at the earliest possible moment - at bug creation - when the available data is minimal. Its purpose is not to deliver a highly accurate forecast, but to act as a low-cost, strategic early warning mechanism. It flags bugs that exhibit characteristics of potential priority instability long before they become critical problems. Practitioners should interpret its output as a signal for increased monitoring, not as a definitive verdict. The trade-off here is accepting a lower level of certainty in exchange for the invaluable advantage of early awareness. The Phase II model, in contrast, prioritizes accuracy. It is triggered later in the bug's lifecycle, only after a rich set of dynamic, evolution features (such as comment activities and fields modification histories) has accumulated. By leveraging this wealth of information, it can provide a much more reliable and accurate prediction of the new priority level. Its value is primarily tactical, providing high-confidence guidance precisely when a re-evaluation is warranted. The trade-off between timeliness and accuracy is sacrificing early prediction of the new priority level for a more dependable, actionable insight of bug priority change at critical moments. Crucially, these two phases are not mutually exclusive but complementary. Phase I acts as a filter, identifying a manageable subset of potentially volatile bugs from a large pool of incoming bugs. Phase II then provides a high-precision ``magnifying glass'' to examine these flagged bugs as they evolve. A comprehensive bug management strategy requires both capabilities.

\textbf{Interpreting the Precision-Recall Trade-off in Practice.} The performance of our models involves a trade-off between Precision and Recall. Understanding the practical implications of this trade-off is essential for developers and managers to tailor the model's application to their specific needs and resource constraints.
\textbf{Optimizing for High Precision (Minimizing False Alarms):}
A model configuration with higher Precision than Recall means that when the model predicts a bug priority change, it is highly likely to be correct. The downside is that it may miss some true priority-change bugs (i.e., false negatives).
For developers, this model is best used as a high-confidence ``Action'' trigger. If a developer receives an alert from this model, he/she should treat it with high urgency, since its predictions are highly reliable with very few false positives. It is ideal for teams with limited resources in manual verification, as it minimizes the time spent on investigating false alarms. 
\textbf{Optimizing for High Recall (Minimizing Missed Risks):}
A model with higher Recall than Precision indicates that the model can identify most of the bugs that are truly going to change their priority. The downside is that it may incorrectly flag a number of bugs with stable priorities as potentially changing.
For developers, this model is best used as a comprehensive ``Monitoring'' or ``Screening'' tool. The alerts it generates should be treated as a list of candidates for a secondary, manual review. It is ideal for risk-averse teams who want to ensure that no potentially critical bugs is missed.
By understanding Precision-Recall trade-off, teams can potentially adjust the model's decision threshold. Lowering the threshold typically increases Recall at the expense of Precision, and vice versa. This allows practitioners to customize the model's behavior based on the team's priority for either minimizing false alarms or preventing missed risks in priority change predictions.

\subsection{Implications for Researchers}
Our exploratory study on bug priority change prediction not only provides a novel method but also opens up several promising avenues for future research. We outline four key implications for researchers.

\textbf{A New Research Direction: From Static Prediction to Dynamic Forecasting.}
Our work marks a clear shift from the well-trodden path of static, initial priority prediction to the more complex and dynamic challenge of priority change prediction. We encourage researchers to move beyond one-time classification at bug report creation and to develop models that can account for the entire bug lifecycle. This focus shift invites a new class of research questions:
Can we predict the number of times a bug's priority will change?
Can we forecast the timing of the next priority change (i.e., time-to-event analysis)?
How do priority change patterns differ across different stages of a project's maturity (e.g., early development vs. maintenance)?

\textbf{Deepening the Understanding of Evolutionary Features.} 
Our findings strongly validate that dynamic, evolutionary features are critical for this prediction task. This opens up a rich area for future feature engineering research. While we have proposed and validated a set of features based on four perspectives, there is significant potential to explore more granular or conceptually different dynamic signals. 
Semantic Analysis of Change Over Time: Instead of just counting comments, future work could analyze the semantic drift in developer conversations over time. For example, does the emergence of terms related to ``performance'' or ``security'' in the comments strongly predict an imminent priority escalation?
Social Network Dynamics: The interaction patterns between developers on a bug report could be modeled as a dynamic graph. Features derived from this graph, such as a sudden increase in the centrality of a key architect, might be powerful predictors.
Fine-grained Code-level Features: Linking bug reports to specific commits and pull requests could allow for the extraction of dynamic code-level features, such as the complexity of the code being modified in response to the bug.

\textbf{Advanced Methodologies for the Priority Change Prediction Task.} 
Our study highlights two critical methodological challenges that warrant further investigation, namely the development of more sophisticated imbalance handling strategies and the exploration of alternative modeling methods. 
Regarding imbalance handling, we demonstrated that the priority change problem suffers from a dual imbalance (rare source classes and skewed transition paths). Our success with conditional sampling over unconditional methods suggests that this is a fruitful direction. Future research should investigate more advanced conditional data generation techniques or cost-sensitive learning algorithms specifically tailored to multi-class problems with imbalanced transition matrices.
As for alternative modeling methods, while our two-phase method is effective, it is not the only possible method. Researchers could explore end-to-end models that jointly predict both if a change will happen and what it will be. Multi-task learning architectures, where one task is the binary prediction and the other is the multi-class prediction, could be a promising alternative. Furthermore, given the sequential nature of bug report events, models explicitly designed for time-series data could be adapted to this problem to better capture the temporal dependencies between events.

\textbf{Additional Exploration of Large Language Models.} Our supplementary exploration with LLMs, specifically GPT-5 and Gemini 2.5 Pro, yields several implications for researchers interested in leveraging such models for software engineering prediction tasks. In Phase I, where only summaries and metadata of 200 bug reports were provided under a zero-shot prompting setting, GPT-5 achieved an accuracy of 0.820 with an F1-score of 0.247, while Gemini 2.5 Pro reached 0.760 accuracy with an F1-score of 0.200. In Phase II, where richer contextual information (including comments and historical records from 100 bug reports) was introduced, GPT-5 attained F1-macro and F1-weighted scores of 0.36 and 0.42, respectively, compared to Gemini 2.5 Pro with 0.39 and 0.36. Notably, few-shot prompting did not provide meaningful improvements over zero-shot prompting in either phase. These results carry three key implications. First, despite their impressive general capabilities, LLMs exhibited only modest effectiveness in predicting priority changes, suggesting that prompt engineering alone is insufficient to capture the subtle decision-making underlying bug priority changes. Second, although the inference costs were relatively modest (ranging from approximately \$0.001 to \$0.009 per prediction), the costs remain non-negligible compared to traditional machine learning and deep learning approaches, which incur nearly zero inference overhead. Finally, since bug report data are publicly available, one cannot exclude the possibility that such data were already included in the pretraining corpora of these LLMs, raising concerns regarding evaluation validity. Taken together, these findings indicate that off-the-shelf LLMs may not be directly suitable for bug priority prediction tasks, and fine-tuning might be a direction worth exploring.

\section{Threats to Validity}\label{sec_validity}

According to the guidelines in \cite{RuHo2009}, we discuss potential threats to the validity of our study following four widely adopted categories: construct validity, internal validity, external validity, and conclusion validity.

\subsection{Construct Validity}

Construct validity refers to the degree to which our study's concepts and measures accurately represent the real-world phenomena they are intended to capture.

\textbf{Ground Truth of Bug Priority Data.} A significant threat to construct validity is that we cannot fully ensure the absolute ground truth of the bug report data. Our study relies on the priority levels assigned and modified by developers in JIRA. We operated under the assumption that developers, particularly those directly involved with a bug, have the most accurate understanding of its priority. However, this may not always be the case. For instance, inexperienced developers might assign incorrect initial priorities or make inaccurate changes. Similarly, developers might occasionally modify priorities for multiple bugs in batches for administrative reasons, which may not reflect a careful, individual re-evaluation of each bug.
To mitigate this threat, we implemented several rigorous data preprocessing steps as described in Section \ref{process_data_processing}. We filtered out bugs with non-standard priority levels to ensure consistency and find batch edits made by the same person at the same time, and then filtered out the priority changes of these batch edits. More importantly, we manually inspected the distribution of bug reports, and established and applied rules to correct or merge multiple rapid priority changes. This helped filter out changes that were likely caused by user errors or immediate regret, rather than intentional changes to incorrect status. Although these measures could not eliminate all potential noise, they significantly improved the quality and reliability of our dataset, ensuring that the data we used is a reasonable and robust representation of actual bug priority dynamics.

\textbf{Conceptualization of Evolution Features.} Our study introduces a novel set of ``bug fixing evolution features'' intended to represent the dynamic changes that occur during a bug's lifecycle. There is a potential threat that these engineered features (e.g., change rate in comment frequency, reporter's historical performance) are imperfect proxies for the complex, real-world bug fixing process. To mitigate this threat, the construction of these features was grounded in findings from prior empirical studies on bug tracking and software maintenance \cite{LiCaYuLiMoLi2024}, as well as a logical understanding of the signals that indicate a bug is receiving increased attention or re-evaluation. For example, a sharp increase in comment activity is a strong and logical indicator that a bug's perceived importance is changing. By validating the predictive power of these features in our ablation studies (RQ1), we provided empirical evidence that they indeed capture meaningful aspects of the bug evolution process.

\subsection{Internal Validity}

Internal validity concerns causal relationships and whether observed outcomes can be attributed to the experimental variables rather than other confounding factors.

\textbf{Data Preprocessing Bias.} In Section \ref{process_data_processing}, we established a 5-minute time limit to correct or merge multiple rapid priority changes made by the same person, aiming to filter out modifications likely caused by user errors or immediate regret. This rule, while based on our manual inspection of the data, is a heuristic and introduces a potential threat. It might have incorrectly filtered out a small number of legitimate, rapid re-evaluations, or failed to capture erroneous changes that occurred just outside this window. We mitigated this threat by choosing a conservative time limit, but we acknowledged that this is a trade-off between data cleanliness and the risk of introducing minor bias.

\textbf{Feature Engineering Choices.} Our study introduces a set of novel bug fixing evolution features. The specific formulas and methods used to calculate these features (e.g., the rate of change in comment frequency) represent a particular design choice. It is possible that alternative calculation methods could lead to different results. This choice presents a potential threat to internal validity. To mitigate this threat, we grounded our feature design in principles from existing literature and a logical understanding of the bug fixing process \cite{AcGi2024, ShAl2022}. More importantly, we conducted extensive ablation experiments (reported in Section \ref{sec_rq1} for RQ1) to systematically evaluate the impact of these features. The results demonstrated that our evolution features significantly improved model performance compared to models using only basic features, thus providing strong evidence for the validity of our choices.

\textbf{Threats Introduced by Sampling Method Selection.} Our choice of K-means undersampling (Phase I) and Conditional Mixed Sampling (Phase II) presents a potential threat to internal validity, as the performance of any sampling method is data-dependent. This introduces the risk that our results could be an artifact of these specific choices. To mitigate this threat, we validated our selections through data-driven pilot experiments. For Phase I, we benchmarked K-means against alternatives including random undersampling, NearMiss (NM), and Edited Nearest Neighbors (ENN). For Phase II, our conditional method was compared against simpler strategies like standard SMOTE and non-conditional mixed sampling. These empirical comparisons demonstrated that our chosen sampling methods are indeed superior to common alternatives for the bug priority change prediction task, which provides strong evidence that the overall performance gains of our model are attributed to its core components (such as our evolutionary features), rather than being a byproduct of an arbitrary or lucky choice of sampling technique.

\textbf{Threats from Model Selection and Hyperparameter Tuning.} Our selection of a specific model ensemble (Random Forest, KNN, SVM, and XGBoost) for Phase I poses a threat to internal validity, as this combination may not be optimal; other models or strategies might yield better results. Additionally, we did not perform exhaustive hyperparameter tuning, relying instead on reasonable configurations from pilot experiments. However, our primary goal was not to build an optimal model but to validate our novel evolution features and two-phase method. The fact that our approach achieves significant gains even without exhaustive optimization provides strong evidence that the performance is driven by our evolution feature design and sampling strategies, rather than by fine-tuning a specific model set.

\subsection{External Validity}

External validity refers to the extent to which our findings can be generalized to other contexts, such as different projects, organizations, or issue tracking systems.

\textbf{Generalizability to Other Issue Tracking Systems.} Our study exclusively used bug report data from the JIRA system utilized by Apache projects. The priority levels (``Blocker'', ``Critical'', etc.) and the structure of bug reports are specific to JIRA. Therefore, our findings may not directly generalize to projects that use other issue tracking systems like Bugzilla, or GitHub Issues, which may have different priority schemes and data fields. However, the fundamental concepts we model - such as reporter history, comments, and change logs - are common across most systems, suggesting that our approach is likely adaptable. Our future work will aim to replicate this study on datasets from other platforms.

\textbf{Generalizability to Other Projects.} We selected 32 large-scale, open-source projects from the Apache Software Foundation (ASF). This poses a threat, as the results may not be generalizable to other types of projects, such as closed-source commercial software, smaller-scale projects, or projects from different domains. The development culture and processes within the ASF are relatively consistent, which could differ from those in other ecosystems. We mitigated this in part by performing a robust cross-project validation (RQ3), which demonstrated that a model trained on a set of ASF projects can still perform decently on another unseen ASF project. This result provides certain confidence in the model's robustness within this specific ecosystem. However, generalizing these findings beyond large-scale open-source Apache projects will require further studies.

\subsection{Conclusion Validity}

Conclusion validity concerns the extent to which conclusions about relationships in our data are reasonable.

\textbf{Reliability of Measures.} A primary threat to conclusion validity is the use of inappropriate statistical measures. We mitigated this threat by carefully selecting evaluation metrics for each phase of our study. For the binary classification in Phase I, we used the standard and well-understood metrics of Precision, Recall, and F1-score. For the multi-class classification in Phase II, the data was highly imbalanced across priority levels. Using simple accuracy would be misleading. Therefore, we chose F1-weighted and F1-macro scores. These metrics provide a more reliable and comprehensive assessment of model performance across all classes, giving appropriate weight to both majority and minority classes.

\textbf{Reliability of Experimental Results.} To ensure our conclusions were robust, we designed our experiments to minimize the risk of random chance. The ablation experiments (RQ1) allowed us to systematically attribute performance improvements to specific components of our method (i.e., evolution features and class imbalance handling). For RQ3, we employed a rigorous cross-project validation methodology (training on N-1 projects and testing on the remaining one), which is a standard way for assessing a model's ability to generalize to new, unseen data. Additionally, we used the Spearman correlation coefficient to statistically confirm the relationship between the probability of priority change and model performance, ensuring that this conclusion was not based on mere observation alone.

\section{Conclusions and Future Work}\label{sec_conclusion}
We proposed a two-phase software bug report priority change prediction method based on bug fixing evolution features and class imbalance handling strategy. 
The bug priority change prediction task was divided into two phases, i.e., the bug reporting phase and the bug fixing phase. Priority change prediction models for these two phases were constructed using ML and DL techniques, respectively. 
First, the results of ablation experiments demonstrated that the bug fixing evolution features and class imbalance handling strategy proposed in this study can effectively improve the performance of the prediction models.  
Second, the results of comparative experiments show that the class imbalance handling strategy adopted in this study is superior to other methods. 
Third, the performance of cross-project prediction models achieves a decent overall level but varies significantly among different projects. 
Finally, the prediction models for different priorities achieve satisfactory and relatively close performance. 
In summary, our proposed method can effectively predict changes in bug priority, helping developers optimize resource allocation and time scheduling, thereby improving the efficiency of the software bug fixing process. 

Future work could focus on the following areas:
(1) Improving the generalizability of the model by training and testing prediction models on bug data of projects from diverse open source software ecosystems.
(2) Predicting the timing of bug priority changes by employing time-to-event analysis. A limitation of our current model is that it does not predict when a change will occur; incorporating such a temporal prediction would enhance its practical value for project planning.
(3) Mining more valuable bug features by exploring additional attributes that can better characterize the dynamic evolution of the bug fixing process, thus improving the performance of the prediction model.
(4) Optimizing the method to handle class imbalance by experimenting with a wider range of algorithms and techniques to enhance the model’s recognition capability for minority classes.

\section*{Data Availability} The replication package of this study has been made available at \cite{dataset}.

\section*{Acknowledgments}
This work was funded by the National Natural Science Foundation of China under Grant Nos. 62176099 and 62172311. The numerical calculations in this paper have been done on the supercomputing system in the Supercomputing Center of Wuhan University.

\bibliographystyle{ACM-Reference-Format}
\bibliography{reference}

\balance

\end{document}